\documentclass[a4paper,11pt]{article}
\pdfoutput=1 

\usepackage{jheppub} 

\usepackage[T1]{fontenc} 

\newcommand\fb{{f_b}}
\newcommand\Kinn{{\bf \Phi}_{\rm\scriptscriptstyle B}}
\def\lq{\left[} 
\def\rq{\right]} 
\def\rg{\right\}} 
\def\lg{\left\{} 
\def\({\left(} 
\def\){\right)} 

\newcommand\sss{%
{\scriptscriptstyle}%
}

\newcommand\scalup{{\tt SCALUP}}
\newcommand\frindsing{{\ctindr}}
\newcommand\ctindr{{\alpha_{\sss\rm r}}}

\newcommand\Rad{\Phi_{\rm rad}}
\newcommand\Kinnpo{{\bf \Phi}_{n+1}}
\newcommand\BKinn{{\bf \bar{\Phi}}_n}
\newcommand\ctindp{{\alpha_{\splus}}}
\newcommand\ctindm{{\alpha_{\sminus}}}
\newcommand\Kinncp{{\bf \Phi}_{n,\splus}}
\newcommand\Kinncm{{\bf \Phi}_{n,\sminus}}
\newcommand\splus{{\sss \nplus}}
\newcommand\sminus{{\sss \nminus}}
\newcommand\nplus{\oplus}
\newcommand\nminus{\ominus}
\newcommand\pt{p_{\sss\rm T}}
\newcommand\pten{(p_\ell+p_\nu)_{\sss\rm T}}
\newcommand\kt{k_{\sss\rm T}}
\newcommand\stepf{\theta}
\newcommand\ptmin{{\pt^{\min}}}
\newcommand\MCFM{\texttt{MCFM}}
\newcommand\POWHEGBOX{\texttt{POWHEG BOX}}
\newcommand\POWHEG{\texttt{POWHEG}}
\newcommand\ALPGEN{\texttt{ALPGEN}}
\newcommand\MADGRAPH{\texttt{MADGRAPH}}
\newcommand\SHERPA{\texttt{SHERPA}}
\newcommand\DIPHOX{\texttt{DIPHOX}}
\newcommand\JETPHOX{\texttt{JETPHOX}}
\newcommand\PYTHIA{\texttt{PYTHIA}}
\newcommand\herwigpp{\texttt{Herwig++}}
\newcommand\PWG{\texttt{PWG}}
\newcommand\FORM{\texttt{FORM}}

\newcommand\mathd{\mathrm{d}}

\newcommand\as{\alpha_{\sss\rm S}}
\newcommand\alphaem{\alpha_{\sss\rm em}}
\newcommand\MHV{M_{\scriptscriptstyle W^*}}

\newcommand\pT{p_{\sss\rm T}}
\newcommand{\tmop}[1]{\ensuremath{{\rm #1}}}
\def\ord#1{{\cal O}\(#1\)}

\title{\boldmath $W\gamma$ production in hadronic collisions using the POWHEG+MiNLO method}

\author[a]{Luca Barz\`e,}
\author[b]{Mauro Chiesa,}
\author[b]{Guido Montagna,}
\author[c]{Paolo Nason,}
\author[d]{Oreste Nicrosini,}
\author[d]{Fulvio Piccinini}
\author[b]{and Valeria Prosperi}

\affiliation[a]{PH-TH Department, CERN, CH-1211 Geneva, Switzerland}
\affiliation[b]{Dipartimento di Fisica, Universit\`a di Pavia, and INFN, Sezione di Pavia, Via A. Bassi 6, 27100 Pavia, Italy}
\affiliation[c]{INFN, Sezione di Milano Bicocca, Piazza della Scienza 3, 20126 Milano, Italy}
\affiliation[d]{INFN, Sezione di Pavia, Via A. Bassi 6, 27100 Pavia, Italy}

\emailAdd{luca.barze@cern.ch}
\emailAdd{mauro.chiesa@pv.infn.it}
\emailAdd{guido.montagna@pv.infn.it}
\emailAdd{paolo.nason@mib.infn.it}
\emailAdd{oreste.nicrosini@pv.infn.it}
\emailAdd{fulvio.piccinini@pv.infn.it}
\emailAdd{valeria.prosperi@pv.infn.it}

\abstract{
We detail a calculation of $W\gamma$ production in hadronic collision,
at Next-to-Leading Order (NLO) QCD interfaced to a shower generator
according to the \POWHEG{} prescription supplemented with the MiNLO
procedure.  The fixed order result is matched to an interleaved
QCD+QED parton shower, in such a way that the contribution arising
from hadron fragmentation into photons is fully modeled.  In general,
our calculation illustrates a new approach to the fully exclusive
simulation of prompt photon production processes accurate at the NLO
level in QCD.  We compare our predictions to those of the NLO program
\MCFM{}, which treats the fragmentation contribution in terms of a photon
fragmentation functions. We also perform comparisons to available LHC
data at 7 TeV, for which we observe good agreement, and provide
phenomenological results for physics studies of the $W\gamma$
production process at the Run II of the LHC. The new tool, which
includes $W$ leptonic decays and the contribution of anomalous gauge
couplings, allows a fully exclusive, hadron-level description of the
$W\gamma$ process, and is publicly available at the repository of the
\POWHEGBOX{}. Our approach can be easily adapted to deal with other
relevant isolated photon production processes in hadronic collisions.
}

\keywords{QCD, Hadronic colliders, Phenomenological models}

\begin{document} 
\maketitle
\flushbottom

\section{Introduction}
\label{sec:intro}

With the discovery of a new scalar particle in the search for the Standard Model (SM) Higgs boson 
by the ATLAS~\cite{Aad:2012tfa} and CMS~\cite{Chatrchyan:2012ufa}  collaborations at the Large Hadron 
Collider (LHC), all the particles postulated by the SM have been identified. 
In parallel to the present efforts which are mainly focused on studying the properties of the newly discovered boson, 
other important studies set the physics agenda of the LHC, ranging from measurements of SM processes 
to the search for new phenomena.

In this general context, diboson production processes play a particularly interesting r\^ole for 
different reasons~\cite{Schott:2014awa}.
They represent the primary backgrounds to Higgs and new physics searches, and provide direct information on the 
self-interactions of the electroweak (EW) gauge bosons. Since the form and strength of the non-abelian gauge couplings are 
fixed by the underlying $SU(2) \otimes U(1)$ symmetry, any deviation of these couplings from their SM values would be 
indicative of new physics.

Aside from $\gamma\gamma$ production, the production processes of a $W$ or $Z$ boson in 
association with an isolated photon
provide the largest and cleanest yields among diboson final states  at hadron colliders, as backgrounds to $W\gamma$ and
$Z\gamma$ production can be significantly reduced through the identification of the $W$ and $Z$ bosons via their leptonic decay modes.
 Measurements of $V\gamma$ ($V = W,Z$) processes from initial analyses at the LHC 
 have been performed both by the ATLAS~\cite{Aad:2014fha,Aad:2013izg,Aad:2012mr,Aad:2011tc} 
 and by the CMS collaboration~\cite{Chatrchyan:2013fya,Chatrchyan:2013nda,Chatrchyan:2011rr}. These measurements have 
 been used to test the SM predictions, to set limits on anomalous triple gauge couplings (ATGCs) and 
 on the production of new vector resonances. Previous measurements of $V\gamma$ final states in hadronic 
 collisions have been made at the 
 Tevatron by the CDF~\cite{Aaltonen:2011zc} and D0~\cite{Abazov:2011qp,Abazov:2011rk} collaborations
 and used to set limits on ATGCs, that are improved by the current analyses at the LHC. Constraints from LEP 
 on ATGCs are summarized in ref.~\cite{Alcaraz:2006mx}. 
 
 At the LHC, the signal events $p p \to \ell \nu \gamma + X, \ell = e, \mu$ (for $W\gamma$ production) and 
 $p p \to \ell^+ \ell^- \gamma + X, \nu \bar{\nu} \gamma + X$ (for $Z\gamma$ production) are modeled in the latest analyses using the 
 leading-order (LO) matrix element generators \ALPGEN~\cite{Mangano:2002ea}, \MADGRAPH~\cite{Alwall:2011uj} and 
 \SHERPA~\cite{Gleisberg:2003xi}. Broadly speaking, the above LO predictions are found to reproduce the shape 
 of the photon distributions and the kinematic properties of the leptons and jets in $V\gamma$
 candidate events. Afterwards, the cross section measurements are compared to the 
 Next-to-Leading Order (NLO) QCD predictions of the 
 parton-level Monte Carlo (MC) program \MCFM{}~\cite{Campbell:2010ff}, that includes the full set of 
 LO diagrams and NLO QCD corrections contributing to $V\gamma$ production, and takes care of 
 the contribution coming from the fragmentation of secondary quarks and gluons into isolated 
 photons via the formalism of (collinear) photon fragmentation functions~\cite{Campbell:2011bn}. The effects 
 of ATGCs can be simulated in \MCFM{} as well.\footnote{Previous calculations of NLO QCD corrections to $V\gamma$ 
 production in hadronic collisions can be found in refs.~\cite{Ohnemus:1992jn,Baur:1993ir,Baur:1997kz,DeFlorian:2000sg}.  
 NNLO QCD corrections to $Z\gamma$ production have been computed in ref.~\cite{Grazzini:2013bna}, while 
 recent progress in the calculation of NNLO QCD corrections to the $W\gamma$ process is documented 
 in ref.~\cite{Grazzini:2014pqa}. NLO QCD 
 corrections to the related processes $W\gamma$ and $Z\gamma$ plus one or two jets have been 
 calculated in ref.~\cite{Campanario:2009um} and 
 refs.~\cite{Campanario:2013eta,Campanario:2014dpa,Campanario:2014wga}, respectively.
NLO EW corrections, not considered in the present study, to $W\gamma$ and $Z\gamma$ production 
at the LHC have been computed in
 the leading-pole approximation in ref.~\cite{Accomando:2005ra} and to $Z\gamma$ production 
 exactly in ref.~\cite{Hollik:2004tm}.}
 Limits on ATGCs are set by ATLAS using \MCFM{} and by CMS 
 using \SHERPA.
 
 The state of the art of the theoretical tools used in the experimental analyses of $V\gamma$ processes at the LHC 
 points out clearly that progress in this area would be welcome. In fact, it is known that 
 LO matrix element generators matched to Parton Showers (PS) provide a reliable description 
 of the shape of the differential cross sections of experimental interest (even in the presence of a high jet multiplicity) but can not predict 
 their normalization with the desired accuracy. On the other hand, the results of 
 NLO parton-level programs must be corrected to compare the predictions to the measured 
 cross sections. Moreover, the lack of higher-order QCD contributions in fixed-order MCs 
 can give rise to biases in the predicted cross sections, especially for those observables
 significantly affected by the contribution of multiple QCD radiation. In particular, in view of the next 
 data taking at the LHC
 at higher energy and higher luminosity, the improvement of the accuracy of the theoretical predictions is 
 becoming a pressing issue, 
 as the experimental errors of the measurements will diminish and work will 
 continue towards highlighting deviations, if any, 
 from the apparent SM behaviour.
 
Given the above motivations, the main aim of the present work is to
provide a new simulation tool for the study of $W\gamma$ production at
the LHC. By doing so, we also detail a new exclusive MC approach to
the simulation of prompt-photon production in hadronic collisions,
which includes a number of novel features with respect to previous
methods proposed in the literature.

Our generator is built according to the \POWHEG{} method~\cite{Nason:2004rx,Frixione:2007vw},
within the \POWHEGBOX{} framework \cite{Alioli:2010xd}, that allows to interface an NLO
calculation to a PS generator. We propose a description of the process
and, in particular, of the fragmentation mechanism which includes the
contribution of higher-order matrix elements interfaced to a mixed
QCD+QED PS. For the treatment of the matrix elements that are not
integrable over the full phase space, as well as to ensure sensible
results and a smooth behaviour near the Sudakov regions, we use the
MiNLO (Multi-scale improved NLO) method developed in
refs.~\cite{Hamilton:2012rf,Hamilton:2012np}. We also include the
contribution of ATGCs according to the standard $CP$-conserving
Lagrangian parameterization adopted in the experimental analyses.

The work presented here is the first NLOPS 
(NLO calculation interfaced to a PS) simulation of $W\gamma$
production in hadronic collisions. Its theoretical framework is novel, and
can be applied to other processes involving the production of isolated
photons. The relative computer code is made available in the public repository of the \POWHEG{}
BOX~\cite{Alioli:2010xd} at the web site {\tt
  http://powhegbox.mib.infn.it}.

The basic idea underlying our method is to treat electromagnetic and
strong radiation on the same footing within the \POWHEG{}
approach. Given the basic process of $q\bar{q}' \to W\gamma$
production, \POWHEG{} generates the strong radiation using the real
matrix elements for the process $j j \to W\gamma j$, where $j$ stands
here for any parton.\footnote{Here and in the following, when we
  indicate a final state $W$ we imply that we are considering its
  leptonic decay.} \POWHEG{} will separate the real cross section
into a sum of different contributions, corresponding to the singular
regions of the real amplitude. If electromagnetic and strong radiation are treated
on the same footing, there will be two kinds of singular
regions: those where the emitted parton $j$ is the collinear one, and
those where the photon is collinear (either to the initial state
partons, or to the electron coming from the decay of the $W$). The two different
kinds of regions will have two different kinds of underlying Born processes,\footnote{The underlying
Born process for a given singular region is obtained by merging the collinear particles
relative to a given singular region.} the $q\bar{q'} \to W\gamma$ ones and a $j j \to W j$ ones.
Thus, in this approach we are forced, for consistency, to introduce also the $j j \to W j$
process as a possible Born process. According to the \POWHEG{} formalism, a $j j \to W j$
initial process may radiate a gluon or a photon, according to competing QED
and QCD Sudakov form factor, with the gluon radiation being favoured. In case
a photon is generated, the event will be passed to a shower generator, that will not
be allowed to produce splittings that are harder than the radiated photon.
The corresponding event will typically have a hard jet, a less hard
photon, and more partons, limited in hardness by the photon hardness. In the more likely
case that a coloured parton is generated, the event will be passed to a shower, that
will not be allowed to produce splittings that are harder than the \POWHEG{} radiated
parton. If the PS generator includes QED radiation, hard photons may also be produced
by the shower. It is clear that in this approach the full photon radiation
phase space will be reconstructed from different components:
\begin{itemize}
\item The  $q\bar{q}' \to W\gamma$ initiated event, where the hardest radiation is a photon
radiation.
\item The  $j j \to W j$ initiated event followed by photon radiation from \POWHEG{},
where the hardest radiation is a coloured parton, and the second hardest is a photon.
\item The $j j \to W j$ initiated event followed by parton radiation
  from \POWHEG{}, where the hardest radiation is a coloured parton,
  the second hardest is also a coloured parton, and where a photon may
  still be generated by the shower as the third, or fourth and so on,
  hardest radiation.
\end{itemize}
Notice that within this approach the direct and fragmentation production
mechanisms are treated in a seamless way. Most photon radiation is treated
perturbatively, either with the LO and NLO matrix elements in \POWHEG{}, or
in the PS within the collinear approximation. Ultimately, the hadronization
step may also lead to photons, and whether or not this transition is treated
correctly will depend upon the degree of accuracy of the shower MC generator.

All major general purpose PS generators implement interleaved QCD and QED 
radiation, thus they model associated photon production 
from a given hard process in the collinear 
approximation (see ref.~\cite{Agashe:2014kda} and references therein).  
An improved approach based upon the usage of LO multiparton matrix elements 
and an interleaved QCD+QED PS can be found in ref.~\cite{Hoeche:2009xc},
where results for the inclusive production of isolated photons and
diphoton production are given in comparison to Tevatron
measurements. In this approach also large angle photon and parton 
emission,  as well as their interplay, is described with LO matrix elements accuracy. 

A first attempt to simulate photon production processes in hadronic
collisions at NLOPS accuracy according to the \POWHEG{} method 
has been developed in ref.~\cite{D'Errico:2011sd} and applied to diphoton
production. We will discuss and clarify similarities and differences
of this method with respect to our schemes, by also providing an optional 
variant of our generator that mimics it closely.

A completely different approach to prompt photon production, applied
to the $t\bar{t}\gamma$ and  $t\bar{t}\gamma\gamma$ production processes, 
has appeared recently in
refs.~\cite{Kardos:2014zba,Kardos:2014pba}. We will further comment about 
this approach in the Conclusions.

The paper is organized as follows. In Sect.~\ref{sec:theory} we describe the theoretical details of our approach, 
paying particular attention 
to the method used for the generation of the hardest emission and for the treatment of the photon 
fragmentation process, which is 
one of the main issues of the calculation. In Sect.~\ref{sec:pheno} we illustrate comparisons of 
our predictions to \MCFM{} calculations and to LHC data at 7 TeV, and we show a sample of 
numerical results for physics studies at the LHC at 14 TeV. We present our conclusions in Sect.~\ref{sec:conc}.

\section{Theoretical framework}
\label{sec:theory}

\subsection{Leading order contributions and anomalous couplings}
\label{sec:lo}

\begin{figure}[t]
\begin{center}
\includegraphics[width=4cm]{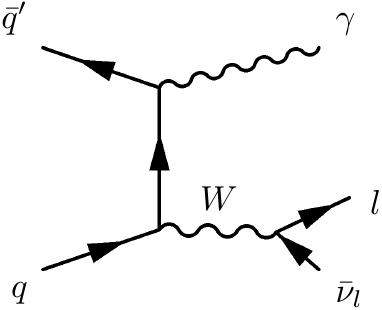}~~~~~~~~~~~~~~~~\includegraphics[width=4cm]{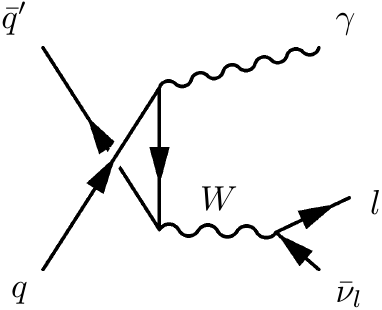}

\includegraphics[width=4cm]{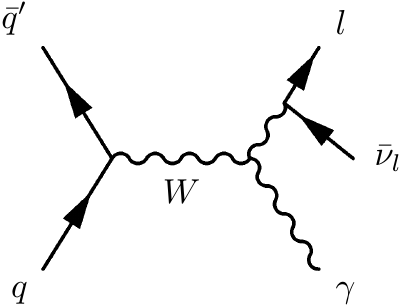}~~~~~~~~~~~~~~~~\includegraphics[width=4cm]{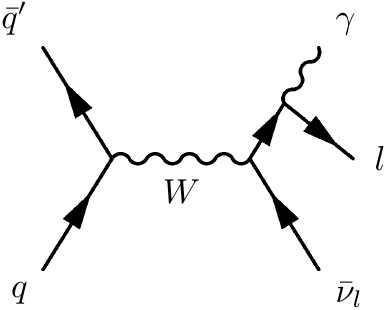}
\end{center}
\caption{\label{LOdiagrams} LO Feynman diagrams for the $\ell \nu\gamma$ production process in hadronic collisions.}
\end{figure}

At LO, the production of a $W$ boson and a photon in hadronic collisions, with leptonic decays of the vector boson,
 is an EW process which 
proceeds via quark-antiquark annihilation
\begin{equation}
q \bar{q}^{\prime} \to \ell^{\pm} \nu \gamma, \quad \ell = e, \mu 
\end{equation}
in terms of the Feynman diagrams shown in Fig.~\ref{LOdiagrams}. The first three diagrams 
are typically classified as direct photon radiation in the production 
process, while the last diagram corresponds to final-state photon emission from the lepton in the $W$ decay. We computed 
the corresponding LO amplitude, which retains full spin correlations in the decay and 
interference effects, by using the computer 
algebra program \FORM{}~\cite{Vermaseren:2000nd}.
A prominent feature of the $W\gamma$ LO matrix element is the appearance of a so called radiation zero, which 
corresponds to the existence of some kinematic configurations for which the amplitude 
vanishes~\cite{Brown:1979ux,Mikaelian:1979nr}. This can 
appear in some observables as dip in the rapidity distributions and can provide a handle to extract information on the
anomalous couplings, since the latter partially fill the dip. It is known that NLO QCD corrections modify the
LO results by partially filling the gap~\cite{Baur:1993ir}. Fortunately, the sensitivity to the 
ATGCs can be largely recovered by imposing a
jet veto~\cite{DeFlorian:2000sg}. Interestingly, an analysis of this kind has been recently made by the 
CMS collaboration~\cite{Chatrchyan:2013fya}
by measuring the differential yield as a function of the charge-signed rapidity difference between a photon candidate 
and a lepton in $W\gamma$ candidates.\footnote{A first study of the radiation-amplitude zero in $W\gamma$ 
production using such an observable was made by the D0 collaboration in proton-antiproton collisions 
at the Tevatron~\cite{Abazov:2008ad}.} The distributions measured at the LHC clearly demonstrate 
the characteristic radiation zero expected for $W\gamma$ production, in agreement with the SM prediction.

The $WW\gamma$ vertex relevant for the limits on ATGCs enters via the third diagram in Fig.~\ref{LOdiagrams}. 
In our calculation, we included the 
contribution of ATGCs according to the parameterization used at the LHC and in previous measurements at hadron and $e^+ e^-$ 
colliders. We 
introduced the anomalous contributions to the $WW\gamma$ vertex in terms of the Feynman rules associated 
to the following effective Lagrangian~\cite{Gaemers:1978hg,Hagiwara:1986vm,Baur:1988qt,Baur:1989gk,Berends:1995dn}
\begin{eqnarray}
{\cal L}_{WW\gamma} \, = \, - i e \left[  \left( W_{\mu\nu}^{\dagger}  W^{\mu} A^{\nu } - W_{\mu}^{\dagger}  A_{\nu} W^{\mu\nu } \right)   
+ k_\gamma W_{\mu}^{\dagger}  W_{\nu} F^{\mu\nu } + \frac{\lambda_\gamma}{M_W^2} W_{\lambda\mu}^{\dagger}  W^{\mu}_{\nu} F^{\nu\lambda}
\right]  \; .
\label{eq:atgc}
\end{eqnarray}
In Eq.~(\ref{eq:atgc}) $A^\mu$ and $W^\mu$ are the photon and $W^-$ field, respectively, 
$W_{\mu\nu} = \partial_{\mu} W_{\nu} -  \partial_{\nu} W_{\mu} $, $F_{\mu\nu} = \partial_{\mu} A_{\nu} -  \partial_{\nu} A_{\mu} $, 
$e$ is the positron charge and $M_W$ represents the $W$ mass. The effective Lagrangian of 
Eq.~(\ref{eq:atgc}) satisfies electromagnetic 
gauge invariance, as well as $C$ and $P$ invariance. In the SM $k_\gamma = 1$ and $\lambda_\gamma = 0$. 
The effect of ATGCs is expressed in terms of their deviation from the SM values, leading to 
the two parameter set $(\lambda_\gamma, \Delta k_\gamma)$, with 
$\Delta k_\gamma \equiv k_\gamma - 1$.\footnote{The theoretical and 
phenomenological drawbacks of the anomalous coupling approach in favor of 
the virtues of a modern effective field theory approach have been recently discussed in 
ref.~\cite{Degrande:2012wf}. However, 
as shown in ref.~\cite{Degrande:2012wf}, the results obtained using the anomalous 
coupling formalism can be reframed
in terms of the effective field theory framework, if only dimension-six operators are considered and 
the anomalous couplings are treated as constants, {\it i.e.} independent of energy.}
The full amplitude resulting from the calculation of the diagrams shown in 
Fig.~\ref{LOdiagrams} with the modifications introduced by the 
Lagrangian of Eq.~(\ref{eq:atgc}) has been computed using \FORM{}.

As a cross-check, we compared our LO predictions, both without and with ATGCs, 
with those of \MCFM{}, finding perfect agreement.

\subsection{NLO QCD corrections}
\label{sec:nlo}

The NLO QCD corrections to $W\gamma$ production can be obtained by dressing the diagrams of Fig.~\ref{LOdiagrams}
with both virtual and real gluon radiation. 

\begin{figure}[t]
\begin{center}
\includegraphics[width=4cm]{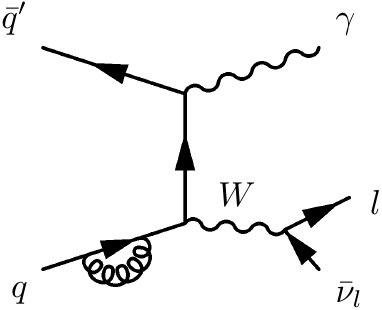}~~~~~~\includegraphics[width=4cm]{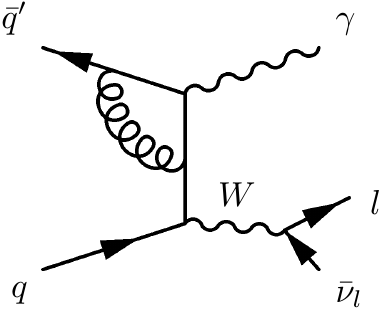}~~~~~~\includegraphics[width=4cm]{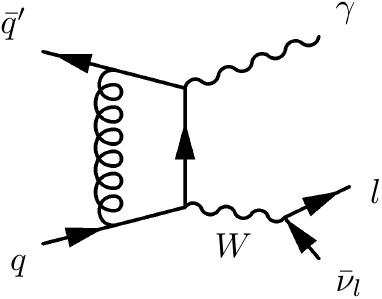}
\end{center}
\caption{\label{Vdiagrams} Sample graphs for the virtual QCD corrections to  $q \bar{q}^{\prime} \to \ell\nu\gamma$ production.}
\end{figure}

The virtual corrections due to the interference of one-loop diagrams with the Born amplitude comprise self-energy, 
vertex and box corrections to the quark lines of Fig.~\ref{LOdiagrams}. Sample graphs for the virtual 
corrections to the $t$-channel  $q \bar{q}^{\prime} \to W\gamma$ topology are shown in Fig.~\ref{Vdiagrams}.

Consistently with the \POWHEG{} method and 
the \POWHEGBOX{} requirements, we  computed the finite part of the virtual cross section in conventional dimensional
regularization, using the Passarino-Veltman tensor reduction \cite{Passarino:1978jh}. In order to compare with the
predictions of \MCFM{}, where the NLO calculation is performed in the dimensional reduction scheme, we translated 
our result for the virtual contribution from dimensional regularization to dimensional reduction, according to 
the rule given in ref.~\cite{Alioli:2010xd} (see also ref.~\cite{Kunszt:1993sd}). We checked that the results of \MCFM{} and 
those of our calculation perfectly agree. However, because the calculation of the one-loop 
corrections using the \MCFM{} matrix elements is less CPU demanding, we included them in our implementation.

\begin{figure}[t]
\begin{center}
\includegraphics[width=5.25cm]{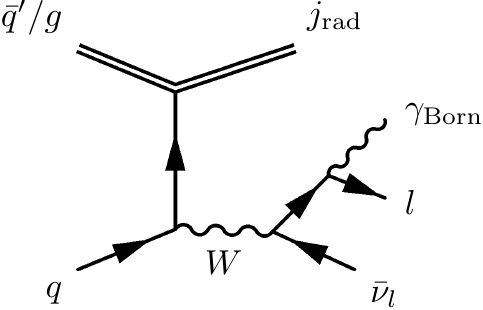}~~~~~~~~~~\includegraphics[width=4cm]{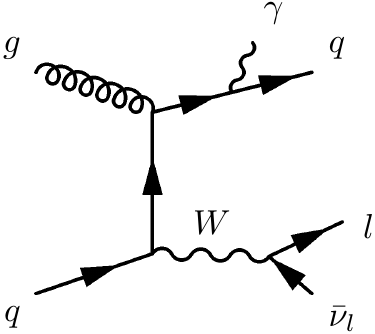}
\end{center}
\caption{\label{Rdiagrams} Sample graphs for real radiation contributions to  $q \bar{q}^{\prime} \to \ell\nu\gamma$ production.
Left diagram: QCD radiation in $\ell\nu\gamma$ production with photon emission from the final-state lepton; right diagram:
gluon-induced process with photon emission from a final-state parton, associated to a fragmentation contribution.}
\end{figure}

The real radiation contributions are obtained by attaching a gluon to the LO diagrams of Fig.~\ref{LOdiagrams} in all 
possible ways. The contributions with one extra parton in the final state are the $2 \to 4$ processes 
$q \bar{q}^{\prime} \to \ell \nu \gamma g$ and $g q/ \bar{q}^{\prime} \to \ell \nu \gamma  \bar{q}^{\prime} / q$. 
Two examples of Feynman diagrams for real radiation contributions to $W\gamma$ production are 
shown in Fig.~\ref{Rdiagrams}.

We computed the real radiation Feynman diagrams using 
\FORM{}. Also in this case, 
we compared our calculation with \MCFM{}, finding perfect agreement. We implemented in the \POWHEGBOX{} 
the real radiation matrix elements extracted from \MCFM{}, as the latter ensure the best CPU performances 
by virtue of the helicity formalism used there.\footnote{Note that 
in the computation of the all real radiation processes yielding the signature $\ell \nu \gamma + j$, $j = q, \bar{q}^{\prime}, g$ 
we included the contribution of ATGCs as well, in order to ensure infrared (IR) cancellation between 
virtual and real corrections in the presence of anomalous couplings.}

The real radiation processes contain QCD singularities due to collinear gluon emission, as well as 
QED singularities corresponding to configurations where a parton becomes collinear to a 
photon, which do not cancel when summing up the real and virtual QCD pieces. 
In the following we will come back to the treatment of this aspect in our approach. 
For the moment, let us remind that in NLO QCD calculations of 
$V\gamma$~\cite{Baur:1993ir,Baur:1997kz,DeFlorian:2000sg}
and other isolated photon production 
processes~\cite{Gehrmann:2013aga,Hartanto:2013aha,Campbell:2014yka} at hadron colliders, 
as well as in fixed-order MC 
programs for prompt-photon production in hadronic collisions, 
like e.g. \MCFM{}, \DIPHOX~\cite{Binoth:1999qq} 
and \JETPHOX~\cite{Catani:2002ny,Aurenche:2006vj}, the QED divergences associated 
to a final-state parton becoming collinear to a photon are treated in terms of
(non-perturbative) quark/gluon fragmentation functions into photons $D_{a}^{\gamma}(z,\mu^2), a = q,g$. 
They  describe
the probability of finding 
a photon with longitudinal momentum fraction $z$ in a quark or gluon jet at a given fragmentation scale 
$\mu$~\cite{LlewellynSmith:1978dc,Bourhis:1997yu}. Since the photon fragmentation functions 
are of the (leading) order $\alphaem / \as$, the 
fragmentation contribution\footnote{This contribution is also known in the literature as ``bremsstrahlung".} is
 of the same order as LO direct production, and at high-energy hadron colliders
can become a relevant source of prompt photon production because of the large 
impact of the gluon distribution function. However, its magnitude strongly depends 
on the applied experimental cuts and can be drastically reduced by imposing appropriate
 isolation criteria. Among the
different criteria for the isolation of photons there are: the cone approach~\cite{Baer:1990ra,Aurenche:1989gv}, 
the democratic approach~\cite{Glover:1993xc} and the smooth isolation prescription~\cite{Frixione:1998jh}. In particular,
according to the latter algorithm, the contribution of the fragmentation mechanism is eliminated by a 
prescription which is IR safe at all orders, and the 
isolated-photon cross section depends on the direct production process only.\footnote{It is worth 
mentioning that a further source of final-state photons comes from the decays of mesons, such 
as $\pi^0, \eta$ or $\omega$. However, such a mechanism, which is much larger than prompt-photon 
production, constitutes a background to high $\pt$ photon production and experimental measurements are
corrected for this effect.} The smooth isolation prescription is widely applied in perturbative QCD calculations
but its implementation is very cumbersome from the experimental point of view since it requires 
ideal detectors with very fine granularity.

The QCD infrared singularities present at intermediate steps of the calculation 
in the real emission and virtual contributions have been treated 
using the FKS subtraction formalism~\cite{Frixione:1995ms} provided by the \POWHEGBOX{}. To regulate 
the singularities due to photon emission in gluon-induced real radiation processes we used the expressions
of the QED counterterms and collinear remnants introduced 
in refs.~\cite{Barze:2012tt,Barze':2013yca} for the implementation 
of the EW corrections to single vector boson production in the \POWHEGBOX{}.

\subsection{Details of the \POWHEG{} implementation}
\label{sec:pow}


In this Section, we describe our method for the treatment of the $W(\to \ell \nu) \gamma$ process.
In the following, for brevity, we will sometimes omit to indicate the $W$ decay product when referring to a process.
The reader should remember that the decay process is always implied.

\subsubsection{The \POWHEG{} method}
\label{sec:powm}

To illustrate our approach, it is first necessary to remind the master formula and the algorithm used in \POWHEG{} 
for the cross section calculation and event generation. 
It is given by (in the standard \POWHEG{} notation)~\cite{Frixione:2007vw}
\begin{eqnarray}
\label{eq:POWHEG}
d\sigma&=&\sum_\fb \bar{B}^\fb(\Kinn)\, d \Kinn
 \Bigg\{ \Delta^\fb\!\(\Kinn,\ptmin\)
\nonumber\\
&+&\!\!\!\!\!\!\sum_{\frindsing\in\{\frindsing|\fb\}}\!\!\!\!\!\!
 \frac{\Big[  d\Rad\;\stepf\(\kt-\ptmin\)
\Delta^\fb\!\(\Kinn,\kt\)\, R\(\Kinnpo\)
\Big]_\frindsing^{\BKinn^\frindsing=\Kinn}}{ B^\fb\!\(\Kinn\)}
\Bigg\} . \phantom{aa}
\end{eqnarray}
For each contributing flavor structure $\fb$ to a given $n$-body process, 
the two basic ingredients of Eq.~(\ref{eq:POWHEG}) are the NLO inclusive cross section
$\bar{B}^\fb(\Kinn)$ and the (modified) Sudakov form factor 
$\Delta^\fb\ \! (\Kinn, \pt)$ for the calculation of the emission probability. According to 
the \POWHEG{} method, the $n$-body configuration is generated according to 
$\bar{B}^\fb(\Kinn)$ and the hardest emission is generated using the 
Sudakov form factor $\Delta^\fb\ \! (\Kinn, \pt)$. Then the events can be showered 
by a PS algorithm ordered in transverse momentum.

The $\bar{B}^\fb(\Kinn)$ cross section at NLO accuracy can be written as
\begin{eqnarray}
\label{eq:NLO}
\bar{B}^\fb(\Kinn) &=& \lq
B\(\Kinn\)+V\(\Kinn\)\rq_\fb
+ \sum_{\frindsing\in\{\frindsing|\fb\}} \int \Big[  d\Rad\,\lg R\(\Kinnpo\) -
 C\(\Kinnpo\)\rg\Big]_\frindsing^{\BKinn^\frindsing=\Kinn}
\nonumber \\
&&+ \sum_{\ctindp\in\{\ctindp|\fb\}} \int \frac{dz}{z}
\,G_\splus^\ctindp\(\Kinncp\)
+ \sum_{\ctindm\in\{\ctindm|\fb\}} \int \frac{dz}{z}
\,G_\sminus^\ctindm\(\Kinncm\) .
\end{eqnarray}

In Eq.~(\ref{eq:NLO}) $B\(\Kinn\)$ is the LO cross section at fixed underlying 
Born flavour $\fb$ and kinematics $\Kinn$, the real contribution $R\(\Kinnpo\)$ is 
the squared matrix element describing the radiation of an extra parton over the radiative 
phase-space variables $d\Rad$, which is regulated by subtracting the 
counterterms $C\(\Kinnpo\)$ containing the same singularities as $R\(\Kinnpo\)$. The 
finite contribution $V\(\Kinn\)$ includes the virtual loop corrections and the counterterms 
integrated over the real emission variables, which cancel the singularities from the
loop corrections. The factors $G_\splus^\ctindp\(\Kinncp\)$ and $G_\sminus^\ctindm\(\Kinncm\)$
are the collinear remnants, that are the finite leftover of the
subtraction procedure that is applied to absorb the initial-state collinear 
singularities into the parton distribution functions (PDFs).

In the \POWHEG{} algorithm, the real contributions are separated
into singular contributions, corresponding to soft and collinear 
emission, labelled by the index $\frindsing$.
Each $\frindsing$ denotes a single flavour structure, and a single
singular region. Each term $R_\frindsing$ is singular only in the
singular region denoted by $\frindsing$. In Eq.~(\ref{eq:POWHEG}) and Eq.~(\ref{eq:NLO}),
the notation $\frindsing\in\{\frindsing|\fb\}$ means all the real
singular contributions that have $\fb$ as underlying Born flavor. The
square brackets with subscript $\frindsing$ and superscript
$\BKinn^\frindsing=\Kinn$ mean that everything inside refers to the
particular real contribution labelled by $\frindsing$, and having
underlying Born kinematics equal to $\Kinn$.

In place of the standard definition of Sudakov form factor based on the usage of 
collinear splitting functions, the modified Sudakov form factor used in \POWHEG{} 
 is defined in terms of the real radiation matrix element
as follows
\begin{equation}
\label{eq:sudakov}
\Delta^\fb(\Kinn,\pt)=\exp\lg
-\sum_{\frindsing\in\{\frindsing|\fb\}}
\int \frac{\Big[  d\Rad\, R\(\Kinnpo\)\, \stepf\(\kt(\Kinnpo)-\pt\)
\Big]_\frindsing^{\BKinn^\frindsing=\Kinn}}{ B^\fb\(\Kinn\)}
\rg\; ,
\end{equation}
where $\kt(\Kinnpo)$ is a function of 
the real phase space that coincides with the transverse momentum of the emitted parton in 
the soft and collinear limit. Given an underlying Born flavour and kinematics
configuration $\fb$, $\Kinn$,
\POWHEG{} generates the first, hardest emission with a probability distribution
equal to the full differential of the Sudakov form factor. In particular, the
transverse momentum of the hardest radiation is generated with a probabilty
distribution proportional to $\mathd \Delta^\fb(\Kinn,\pt) $.
This is achieved in practice, by writing the Sudakov form factor of Eq.~(\ref{eq:sudakov}) 
as a product of individual Sudakov form factors associated with each $\frindsing$,
generating one $\pt$ value for each one of them, and picking the largest one,
according to the so called ``highest bid method''. Thus, the $\frindsing$
regions compete among each other for the generation of radiation.
Following the hardest radiation, subsequent radiations are simulated
via a PS, with the restriction that radiation harder than the \POWHEG{} generated one is
vetoed.

\subsubsection{Treatment of the direct photon and photon fragmentation contribution}
\label{sec:he}
Here we detail how the various components of our calculation have been included 
in the \POWHEGBOX{}, and how the \POWHEG{} method has been adapted to deal with the direct photon 
and photon fragmentation contributions.

We realized two implementations of the $W\gamma$ process, that differ
in the treatment of radiation for events with $W j$ underlying Born.
We label them as NC (that stands for ``\emph{with no competition}'')
 and C (``\emph{with competition}''), since the difference is related to whether or not
in the generation of radiation in events with $W j$ underlying Born, the
parton emission competes with photon emission, as we will clarify in the following.
The NC implementation uses a method very similar to the one proposed by D'Errico and Richardson
in ref.~\cite{D'Errico:2011sd}, and we include it to make contact with the approach proposed there,
and to clarify the differences with the C implementation, that is the one that we advocate.

In both the NC and C scheme, the Born subprocesses are those for $q\bar{q}'\to \ell \nu \gamma$ and the
$q\bar{q}'\to \ell \nu g$ subprocess with its crossings, corresponding to an incoming gluon and quark
or antiquark.

In the NC scheme, the real subprocesses are all the $j j \to \ell \nu \gamma j$ processes, where $j$
stands for any parton. The  \POWHEGBOX{} separates automatically all singular regions of the real
subprocesses. The regions characterized by a collinear parton $j$ have an associated
underlying Born with an $\ell \nu \gamma$ final state, while those characterized by a collinear photon
have an underlying Born with an $\ell \nu j$ final state.

In the computation of the $\bar{B}$ function for the $q\bar{q}'\to \ell \nu \gamma$ subprocess, we include
the strong soft-virtual corrections (the $V$ term), the collinear remnants and the real contribution
corresponding to a coloured parton becoming collinear in the  $j j \to \ell \nu \gamma j$ real process.
In the case of the $j j \to \ell \nu j $ underlying Born, the $V$ term and the collinear remnants term, corresponding
to electromagnetic corrections are set to zero. The real contribution, corresponding to the collinear photon
region of the  $j j \to \ell \nu \gamma j$ real process (also corresponding to an
electromagnetic correction) is instead included.\footnote{The \POWHEGBOX{} includes it automatically. 
However, excluding them (from the ${\bar B}$ function)
completely would not spoil the accuracy of our calculation, since other corrections of the same 
order ({\it i.e.} the $V$ term) are not included.}
In a variant of the C scheme that will be described later, all strong corrections to the $j j \to \ell \nu j$ 
underlying Born kinematics are also included.

As already recalled, in \POWHEG{} the hardest radiation is generated through the modified Sudakov form factor, by evaluating 
the emission probability for all the allowed IR singular regions. In typical applications, this amounts to considering radiation
from each coloured leg.
The \POWHEGBOX{} can be optionally instructed to also consider the singular regions arising from
electromagnetic radiation, and in the case at hand we turn on this option. Thus IR 
singularities can originate from QCD radiation from partons, as well as from QED radiation off 
quarks and final-state leptons. This situation is somehow similar to what happens in the 
combined treatment of QCD and EW corrections to a 
given hadroproduction process, like e.g. the single $W/Z$ production processes addressed 
in refs.~\cite{Barze:2012tt,Barze':2013yca}. 
However, here the situation is much more complex because of the presence of two inequivalent underlying
Born structures that refer to two different physical processes. If the singular region shows up in 
correspondence with a 
QCD radiation process, it will be driven by gluon 
bremsstrahlung or $g \to q \bar{q}$ collinear splitting in $W\gamma$ production, whose
underlying Born structure is the direct photon contribution. 
On the other hand, when the IR singular configuration comes from an enhanced photon emission off partons/leptons, 
it will be originated by QED emission starting from an underlying $Wj$ Born structure.

In \POWHEG{}, the underlying Born kinematics and flavour is generated first with a probability
proportional to the ${\bar B}$ function. Depending upon this choice, a coloured parton radiation
or a photon radiation is generated at the subsequent stage.
More precisely, our NC scheme is codified in the 
following \POWHEG{} formula:
\begin{eqnarray}
\label{eq:POWHEG-first}
d\sigma&=&\sum_\fb \bar{B}^\fb_{W\gamma}(\Kinn)\, d \Kinn
 \Bigg\{ \Delta^\fb \!\(\Kinn,\ptmin\)
\nonumber\\
&+&\!\!\!\!\!\!\sum_{\frindsing\in\{\frindsing|\fb\}}\!\!\!\!\!\!
 \frac{\Big[  d\Rad\;\stepf\(\kt-\ptmin\)
\Delta^\fb \!\(\Kinn,\kt\)\, R_{W\gamma;j} \(\Kinnpo\)
\Big]_\frindsing^{\BKinn^\frindsing \!\!\!\! = \,\Kinn}}{ B^\fb_{W\gamma}\!\(\Kinn\)}
\Bigg\}  \nonumber\\ 
&+& \sum_\fb {B'}^\fb_{Wj}(\Kinn)\, d \Kinn
 \Bigg\{ \Delta^\fb \!\(\Kinn,\ptmin\) \nonumber\\
 &+&\!\!\!\!\!\!\sum_{\frindsing\in\{\frindsing|\fb\}}\!\!\!\!\!\!
 \frac{\Big[  d\Rad\;\stepf\(\kt-\ptmin\)
\Delta^\fb \!\(\Kinn,\kt\)\, R_{Wj;\gamma} \(\Kinnpo\)
\Big]_\frindsing^{\BKinn^\frindsing \!\!\!\! = \, \Kinn}}{ B^\fb_{Wj}\!\(\Kinn\)}
\Bigg\}  \;,
\end{eqnarray}
where we use the notation $R_{W\gamma;j}^\frindsing$/$R_{Wj;\gamma}^\frindsing$ to denote the contributions to
$R_{W\gamma j}$ that are singular only when a parton/photon is collinear.
Thus, the first two lines of Eq.~(\ref{eq:POWHEG-first}) are associated to the direct photon contribution
({\it i.e.} to the $W\gamma$ underlying Born)
and the last two lines refer to radiative photon contribution (with the $Wj$ underlying Born).
Note that the $R_{W\gamma;j}$ term has only one singular region, corresponding to a radiated parton
collinear to the beam axis, while the $R_{Wj;\gamma}$ has two singular regions, one corresponding to
a radiated photon collinear to the beam axis, and the other corresponding to a photon collinear to
the lepton.

In the $\bar{B}^\fb_{W\gamma}$ term the NLO QCD corrections to the inclusive cross section 
of the direct photon production process are included, while ${B'}^\fb_{Wj}$ has a formal 
structure similar to Eq.~(\ref{eq:NLO}) but with the (QED-like) finite part $V\(\Kinn\)$
(and collinear remnants) set to zero.
For the remaning QED terms we employ the ingredients already available 
in the \POWHEGBOX{} for the treatment of the EW corrections to single 
vector boson production~\cite{Barze:2012tt,Barze':2013yca}.
Therefore, we do not take into account the full NLO EW corrections to the $W j$ process. These corrections
are of $\alphaem$ relative order, and are therefore
subleading. In all cases, they are certainly much smaller than the strong corrections to the $W j$ process.

\subsubsection{Implementation of the NC scheme}

The NC realization thus proceeds according to the following algorithm automated in the \POWHEGBOX{}:
\begin{enumerate}
\item generate an underlying Born kinematics according to the probability distribution
\begin{equation}
\mathd \Phi_{\rm B}{\bar B}_{\rm tot} =\mathd \Phi_{\rm B} \left({\bar B}^\fb_{W\gamma} + {B'}^\fb_{Wj} \right)\;,
\end{equation}
then select a direct photon production or a radiative photon contribution with probability proportional to
${\bar B}^\fb_{W\gamma}(\Phi_{\rm B})$ and 
${B'}^\fb_{Wj}(\Phi_{\rm B})$;
\item once one of the two underlying Born process has been selected, 
generate the hardest radiation using the
corresponding Sudakov form factor. Observe that in case of the 
$W \gamma$ underlying Born the Sudakov form factor refers to QCD emission 
({\it i.e.} to the $R_{W\gamma; j}$ emission), while for the 
$W j$ underlying Born it refers to QED emission 
({\it i.e.} to the $R_{W j; \gamma}$ emission);
\item  in case of events from ${\bar B}^\fb_{W\gamma}$, proceed as in the 
default \POWHEG{} method:
the variable \scalup{} is set to the transverse momentum of the radiated parton, or to
$\pt^{\rm min}$ if no radiation occurs, and the event is passed to the shower
generator;
\item in case of events from ${B'}^\fb_{Wj}$, set \scalup{} to the transverse momentum
of the parton in the underlying Born process, and pass the event to the shower
(notice that the default \POWHEG{} behaviour would instead set \scalup{} to the transverse momentum
of the photon).
\end{enumerate}
In the NC procedure, a photon is always generated at the \POWHEG{} level. It
is thus not necessary to turn on QED radiation in the Shower generator.
If QED radiation is turned on in the Shower, care must be taken to veto
QED radiation harder than \scalup{}, as detailed in the following.

It is useful to see how the full phase space for photon radiation is generated,
without overcounting with this procedure:
\begin{itemize}
\item
in case of events from a $\bar{B}^\fb_{W\gamma}$ underlying Born, we generate events where
{\bf the photon is harder than any other parton} ({\it i.e.} jet) in the event.
By hardness we mean here the $\pt$ relative to
all other particles that could have emitted the photon or parton, including the incoming ones.
One further parton, softer than the photon, is (generally) generated by \POWHEG{}.
The value of \scalup{} is set to the transverse momentum of this radiation, and the shower
generates radiation softer than \scalup{};
\item 
in case of events from ${B'}^\fb_{Wj}$, an event is generated first with the
{\bf the photon softer than at least one emitted parton}.
Since \scalup{} is set in this case to the transverse momentum of the parton
in the underlying Born kinematics, and the photon is softer than this parton, the shower
may still generate parton that are harder than the photon, but softer than the initial,
underlying Born parton.
\end{itemize}
Notice that if we had followed the standard \POWHEG{} procedure for setting \scalup{},
we would have ended up with events where the photon is the second hardest parton with a probability
that is not suppressed by $\alphaem$. In fact, the underlying Born cross section for
the $Wj$ process does not carry a QED coupling, and the Sudakov mechanism for photon emission
guarantees that one photon is always emitted (or at least it would do so if the IR cutoff
for photon emission was set to zero), and no parton harder than the photon could be produced.
This is clearly unphysical. If we suppress parton radiation, a corresponding Sudakov form factor
should also be present, and we don't have it in this case. This is why we must allow harder parton radiation,
by a different setting of the \scalup{} value.

As mentioned earlier, the NC approach to the NLOPS simulation of the $W\gamma$ process closely resembles 
the \POWHEG{}-like method developed in ref.~\cite{D'Errico:2011sd} for the diphoton production process.
That work is carried out in the framework of \herwigpp{}~\cite{Bahr:2008pv}, and truncated shower are provided there
to cope more accurately with the shower matching needed in the case of angular ordered parton showers.
Also in ref.~\cite{D'Errico:2011sd}, events with one less photon and an extra parton ({\it i.e.} 
the $j\gamma$ Born subprocess for the $\gamma\gamma$ final state) are allowed to shower
from the initial scale, using truncated showers, vetoing QED radiation harder than the generated photon,
but imposing no veto on the QCD radiation. Our NC approach is thus equivalent, since truncated showers
are not needed in the $\pt$ ordered showers 
generators \PYTHIA{} v.6 and v.8~\cite{Sjostrand:2006za,Sjostrand:2007gs}
that we are using.

\subsubsection{Implementation of the C scheme}

In the NC scheme, in the generation
of  $Wj$ events, while the QED radiation is emitted using exact tree level
matrix elements, QCD radiation harder than the photon 
is emitted by the shower in the collinear approximation.
This is bound to spoil the accuracy of the QED emission matrix element, since harder QCD radiation
may throw off shell the propagator of the photon-emitting parton by a larger amount than the corresponding QED
radiation.
We thus consider a more accurate description, corresponding to our C scheme, 
codified by the following formula
\begin{eqnarray}
\label{eq:POWHEG-second}
d\sigma&=&\sum_\fb \bar{B}^\fb_{W\gamma}(\Kinn)\, d \Kinn
 \Bigg\{ \Delta^\fb \!\(\Kinn,\ptmin\)
\nonumber\\
&+&\!\!\!\!\!\!\sum_{\frindsing\in\{\frindsing|\fb\}}\!\!\!\!\!\!
 \frac{\Big[  d\Rad\;\stepf\(\kt-\ptmin\)
\Delta^\fb \!\(\Kinn,\kt\)\, R_{W\gamma;j} \(\Kinnpo\)
\Big]_\frindsing^{\BKinn^\frindsing \!\!\!\! = \,\Kinn}}{ B^\fb_{W\gamma}\!\(\Kinn\)}
\Bigg\}  \nonumber\\ 
&+& \sum_\fb {B'}^\fb_{Wj}(\Kinn)\, d \Kinn
 \Bigg\{ \Delta^\fb \!\(\Kinn,\ptmin\) \nonumber\\
 &+&\!\!\!\!\!\!\sum_{\frindsing\in\{\frindsing|\fb\}}\!\!\!\!\!\!
 \frac{\Big[  d\Rad\;\stepf\(\kt-\ptmin\)
\Delta^\fb \!\(\Kinn,\kt\)\, R_{Wj;\gamma} \(\Kinnpo\)
\Big]_\frindsing^{\BKinn^\frindsing \!\!\!\! = \, \Kinn}}{ B^\fb_{Wj}\!\(\Kinn\)} \nonumber \\
 &+&\!\!\!\!\!\!\sum_{\frindsing\in\{\frindsing|\fb\}}\!\!\!\!\!\!
 \frac{\Big[  d\Rad\;\stepf\(\kt-\ptmin\)
\Delta^\fb \!\(\Kinn,\kt\)\, R_{Wj;j} \(\Kinnpo\)
\Big]_\frindsing^{\BKinn^\frindsing \!\!\!\! = \, \Kinn}}{ B^\fb_{Wj}\!\(\Kinn\)}
\Bigg\}  \;.
\end{eqnarray}
In the second term of Eq.~(\ref{eq:POWHEG-second}) 
the Sudakov form factor has now the form 
\begin{eqnarray}
\label{eq:sudakov-C}
&&\Delta^\fb(\Kinn,\pt) = 
\exp\lg
-\sum_{\frindsing\in\{\frindsing|\fb\}}
\int \frac{\Big[  d\Rad\, R_{W j;\gamma}\(\Kinnpo\)  
\, \stepf\(\kt(\Kinnpo)-\pt\)
\Big]_\frindsing^{\BKinn^\frindsing=\Kinn}}{ B^\fb\(\Kinn\)}
\rg \nonumber \\
&&\quad \quad \times
\exp\lg
-\sum_{\frindsing\in\{\frindsing|\fb\}}
\int \frac{\Big[  d\Rad\,  R_{W j; j}\(\Kinnpo\) 
\, \stepf\(\kt(\Kinnpo)-\pt\)
\Big]_\frindsing^{\BKinn^\frindsing=\Kinn}}{ B^\fb\(\Kinn\)}
\rg\; . 
\end{eqnarray}
Thus, in case of the $W j$ underlying Born, QCD radiation competes 
with QED radiation. Because of the larger value of the QCD coupling constant 
QCD radiation will be more frequent. In case of QCD emission from $Wj$ dynamics, 
the photon may only emerge from the subsequent PS, 
where interleaved QED evolution must be turned on. 
The recipe for the C scheme is thus as follows: 
\begin{enumerate}
\item generate an underlying Born kinematics according to the probability distribution
\begin{equation}
\mathd \Phi_{\rm B}{\bar B}_{\rm tot} =\mathd \Phi_{\rm B} \left({\bar B}^\fb_{W\gamma} + {B'}^\fb_{Wj} \right)\;,
\end{equation}
then select a direct photon production or a radiative photon contribution with probability proportional to
${\bar B}^\fb_{W\gamma}(\Phi_{\rm B})$ and 
${B'}^\fb_{Wj}(\Phi_{\rm B})$;
\item if the $W\gamma$ case is selected, radiation is performed 
(as in the NC scheme) according to the QCD Sudakov form factor for 
the emission of an additional parton. The subsequent PS is vetoed 
according to the standard \POWHEG{} \scalup{} value, set 
to the $\pt$ of the emitted parton. Since in this case the PS 
must be used with interleaved QED evolution turned on, care must 
be taken to veto photon radiation with transverse momenta above 
\scalup{};
\item if the $Wj$ case is selected, both QED or QCD radiation can 
be generated. In this way, photon emission off partons turns out 
to be inhibited by the competing QCD radiation. In case of QCD 
emission the photon can only arise from the subsequent PS evolution,
that must be fully turned on, with electromagnetic radiation that
can arise from the final state lepton and from the incoming and outgoing quarks.
Also in this case \scalup{} is set according to the standard 
\POWHEG{} prescription both for photon and parton radiation, 
and care must be taken to veto photon emission (generated by the PS) 
harder than \scalup{}; 
\item
concerning the accuracy
of the $Wj$ contribution, 
we adopt in our formulation two possible options: {\it i)} a LO 
accuracy of the $W$ plus one jet process;  {\it ii)} a NLO accuracy,
obtained by the inclusion of the full, NLO accurate $\bar{B}^\fb_{Wj}$
instead of ${B'}^\fb_{Wj}$. The difference between the 
results of these two variants will be shown and discussed in Section~\ref{sec:pheno}.
We will label the two variants as C-LO and C-NLO.
All the  above improvements were easily implemented using
the routines already available in the \POWHEGBOX{} for the simulation of 
the vector boson plus one jet production process~\cite{Alioli:2010qp}.
\end{enumerate}
A clear representation of the role of the various contributions adopted 
in our C scheme is given below:
\begin{itemize}
\item
Contributions with gluon radiation from a $W\gamma$ underlying Born configuration:
these correspond to the case when {\bf the hardest particle among all
final state partons and photons is a photon}.
\item
Contributions with a photon radiated from a $W j$ underlying Born configuration:
these correspond to the case when {\bf the hardest particle among all
final state partons and photons is a coloured parton; the second hardest one is a photon}.
\item
Contributions with a parton radiated from a $W j$ underlying Born configuration:
these correspond to the case when {\bf the first two hardest particles among all
final state partons and photons are coloured partons}.
In this sample, the shower has to generate correctly the cases when the photon is the third,
fourth, and so on, hardest particle, up to the case when no photon is emitted at all.
\end{itemize}
It is clear now that in order to correctly describe all the classes of events corresponding to the
third item above,  the usage of an interleaved QCD+QED PS is mandatory. 

We remark that the C scheme differs from the NC scheme (and thus also 
from the approach of D'Errico and Richardson~\cite{D'Errico:2011sd}) by the treatment of 
the second parton radiation. In the C scheme, when the second parton 
radiation is harder than the photon radiation, the photon is generated 
by the PS (in the collinear approximation) while the parton is 
generated with matrix element accuracy. Conversely, in the NC scheme 
it can happen that a second parton harder than the photon is emitted 
by the PS in the collinear approximation, while the photon was accurate 
at the matrix element level. This is clearly incorrect, since in this 
case the momentum of the radiated second parton would heavily 
affect the matrix element for photon radiation. 

The C scheme is our novel proposal for the treatment of a prompt-photon process.
This scheme guarantees that photon emission is treated consistently even if the photon
is softer than a certain number of QCD partons. We will see in the following that
the emission of a photon softer than other QCD partons in the event can give sizeable
contributions to realistic observables, and thus a consistent treatment of these
events leads to an improved description of the process.

Our C-NLO scheme, based upon the usage of the $Wjj$ matrix elements and NLO 
accuracy of the $Wj$ process interfaced to a QCD+QED interleaved shower,
goes beyond the required accuracy of our generator, and goes in fact
in the direction of a calculation of the $W\gamma$ process at NNLO accuracy. Nevertheless,
in view of the large NLO corrections to the $Wj$ process, and in view of the fact that
this subprocess contributes to a precise slice of phase space of the whole process
({\it i.e.} no other subprocess can accidentally cancel its contribution), we believe that
the inclusion of its NLO corrections is justified.
We also notice that, formally, when computing NLO corrections to prompt photon production
processes at fixed order using the fragmentation function formalism, the fragmentation component
should also be evaluated at the NLO level. In fact, the fragmentation function is of order
$\alphaem/\as$, so that at the Born level the direct and fragmentation components are
formally of the same accuracy.

\subsection{The MiNLO procedure}
\label{sec:minlo}

As described in the previous Sections, the simulation of  the $W\gamma$ signature rests on the 
calculation of LO and NLO  matrix elements describing $W$ production in association with
real photon and jet radiation. These radiation processes contain singularities associated
with the emission of soft and collinear partons or photons which, in the 
absence of the virtual counterpart, prevent an integration of the related matrix elements over the full phase space. 

The LO $W\gamma$ matrix element is characterized by the presence of singularities associated with
the emission of soft and collinear photons off the partons and off the final state lepton from $W$ decay. This 
requires the introduction of generation cuts that prevent the singular regions from being probed, that must be 
chosen much smaller than the cuts applied at the analysis level. 
In analogy to the \POWHEG{} treatment of the $Vj$ process~\cite{Alioli:2010qp},
we require in the simulation of the $W\gamma$ contribution
the presence of a generation cut of order $\pt^{\gamma, {\rm min}} = 1$~GeV and 
$\Delta R_{\ell\gamma}^{\rm min} = 0.1$, that are definitely smaller than the typical cuts imposed on the 
isolated photon, generally
required to be sufficiently hard (with $\pt^{\gamma} \geq$~10 GeV), and well separated 
from the lepton. This simple procedure guarantees that the results of 
the calculation are stable against variations of the applied experimental cuts
in the case of realistic event selections. Of course, this treatment of the phase space 
turns out to be effective and does not introduce any bias in the theoretical predictions
since the generation cuts
are imposed in a sufficiently loose way on the same particle (the photon) which is at the end more strongly
constrained at the analysis level.

In our approach to $W\gamma$ production, however, a problem arises due
to the partitioning of the final state phase space that we adopt. Let
us consider photons arising from $W$ decays. They will mostly have a
small relative $\pt$ of the photon-lepton system. Events of this kind
arising from the $W\gamma$ underlying Born will be treated by
\POWHEG{} as small $\pt$ events, and will thus have a small \scalup{}
value, such that not much further radiation will be allowed in the
shower. On the other hand, events of the same kind may also arise with
$W j$ as underlying Born.  In the limit of small transverse momentum
of the emitted jet, the divergent underlying Born will end up giving a
sizeable contribution to this kind of events.  In fact, as the
transverse momentum of the underlying Born jet is reduced, the
suppression of photon radiation (due to the reduction of the region where the
photon-lepton $\pt$ is smaller than the jet $\pt$) will compensate the enhancement
of the underlying Born, leaving a significant contribution.  Notice that
photon radiation off quarks will be
irrelevant in this case, since it will not pass the requirement of a hard photon.
On the other hand, radiation from the lepton may still be capable to
yield photons with a hard $\pt$ with respect to the beam axis.
If we require a minimum $\Delta R$ separation between the
photon and the lepton, a small relative transverse $\pt$ of the lepton-photon
system will only be possible if the lepton is soft (we don't consider a soft
$\gamma$, since that will not pass our cuts). We find that, even in this case,
the phase space suppression of this region is not sufficient to fully compensate
the diverging underlying Born cross section, leaving a finite contribution
characterized by a hard photon and a soft lepton, that strongly depends upon the
generation cut for the $W j$ kinematics.

In order to tackle this problem, and to provide a generator able to give predictions under 
general event selection conditions, we resort to the MiNLO procedure~\cite{Hamilton:2012np,Hamilton:2012rf}.
MiNLO can be seen as an NLO extension of the matrix element reweigthing
method used in tree-level matrix element-PS merging
algorithms~\cite{Catani:2001cc,Lonnblad:2001iq,Krauss:2002up, Mrenna:2003if}. 
It has been already applied to the simulation of Higgs and vector boson production
in association with up to two jets~\cite{Hamilton:2012np,Campbell:2013vha} and 
to $HW/HZ$+1 jet~\cite{Luisoni:2013cuh}.
In the MiNLO method, the calculation of an inclusive cross section is modified by the inclusion of Sudakov form 
factors and by making appropriate choices for the scales of the the strong coupling constants associated
with each emission vertex.

In our case, we apply the MiNLO procedure
in analogy to the case of the $Vj$ generator~\cite{Hamilton:2012np}. However, we need 
to specify a slightly different  procedure for the treatment of the $W\gamma$ contribution.

For $Wj$ production at NLO (as in our C-NLO scheme) the application of the
MiNLO procedure is exactly the same as in~\cite{Hamilton:2012np}. 
For $Wj$ at LO, ${B'}_{Wj}$ is modified according to the following
formula
\begin{eqnarray}
{B'}_{Wj} &=& B_{Wj}+\int \mathd \Phi_{\rm rad}   R_{Wj;\gamma} \nonumber \\
&\rightarrow&  B_{Wj} \times \frac{\as \left( \pT \right)}{\as} \Delta^2 \!\left( \MHV ,
  \pT \right) +\int \mathd \Phi_{\rm rad} \frac{\as \left( \pT \right)}{\as} \Delta^2 \!\left( \MHV ,
  \pT \right)  R_{Wj;\gamma},
\label{eq:sudagen}
\end{eqnarray}
where $\MHV$ is the virtuality of the $W$ boson, {\it i.e.} $\MHV^2 = (p_\ell + p_\nu)^2$, and $\pT = \pten$,
and $\as$ in the denominator is evaluated at the same scale as in the $B_{Wj}$ term (in other words,
the $\as$ coupling in  $B_{Wj}$ is replaced by $\as \left( \pT \right)$).
A modification with the same formal structure as in Eq.~(\ref{eq:sudagen})
is applied in the calculation of 
the real radiation matrix element of the $Wj\gamma$ production process, where $\pT$ is the 
transverse component of $(p_\ell + p_\nu)$ computed according to the real radiation kinematics. 
In Eq.~(\ref{eq:sudagen}) the Sudakov 
form factor $\Delta$ is given by 
\begin{equation}
\label{eq:gluon_sudakov}
  \Delta \left( Q, \pT \right) = \exp \left\{ - \int_{\pT^2}^{Q^2}
  \frac{d q^2}{q^2} \left[ A \left( \as \left( q^2 \right) \right) \log
    \frac{Q^2}{q^2} + B \left( \as \left( q^2 \right) \right) \right]
  \right\},
\end{equation}
where the functions $A$ and $B$ have a perturbative expansion in terms of constant coefficients
\begin{equation}
\label{eq:A_and_B}
A \left( \as \right) = \sum_{i = 1}^{\infty} A_i \, \as^i\,, \hspace{2em} B \left(
\as \right) = \sum_{i = 1}^{\infty} B_i \,\as^i \,.
\end{equation}
In MiNLO only the coefficients $A_1, A_2, B_1$ and $B_2$ are needed 
and their expression can be found in ref.~\cite{Hamilton:2012np}.  
At the NLO accuracy, the $Wj$ 
inclusive cross section is treated according to the formula
\begin{eqnarray}
\label{eq:sudagenwjnlo}
\bar{B}_{Wj}& \longrightarrow & \frac{\as(\pT)}{\as}  \Delta^2(\MHV,\pT) \left[ B_{Wj}
\left( 1 - 2 \Delta^{(1)}\right) + \frac{\as(\pT)}{\as} \left(V
+ \int \frac{\mathd z}{z} G \right)\right]  \nonumber\\
 &+&  \int \mathd \Phi_{\tmop{rad}} \frac{\as(\pT)}{\as}
\Delta^2 (\MHV,\pT) \left[\frac{\as(\pT)}{\as}  R_{Wjj}+R_{Wj;\gamma}\right], 
\end{eqnarray}
where $\MHV$ and $\pT$ have the same meaning as in Eq.~(\ref{eq:sudagen}), again with the only difference that
for the real radiation
matrix element $R \equiv |M_{Wjj}|^2$ the variable $\pT$ is derived according to the real radiation kinematics. 
In Eq.~(\ref{eq:sudagenwjnlo})
 $\Delta^{\left( 1 \right)}$ is the $\ord{\as}$ expansion of the Sudakov form factor
\begin{equation}
 \Delta^{\left( 1 \right)}\!\left( Q, \pT \right)= \as (\pt) \left[ 
-\frac{1}{2} A_1 \log^2\frac{\pt^2}{Q^2} + B_1 \log\frac{\pt^2}{Q^2} \right] ,
\end{equation}
which is subtracted in order to maintain NLO accuracy.

Concerning the $W\gamma$ direct photon contribution, we proceed as follows. 
Since the $W\gamma$ contribution can be regarded as a configuration where no parton is emitted 
with transverse momentum 
larger than that of the photon, the photon $\pt$ provides an upper limit for the partonic emission, 
and the proper  reweighting for the LO inclusive cross  $B_{W\gamma}$ is
\begin{equation}
\label{eq:sudagenw}
B_{W\gamma} \longrightarrow B_{W\gamma} \times  \Delta^2 \!\left( \MHV ,
  \pT \right) ,
\end{equation}
where we choose the scales $\pt$ and $\MHV$ taking care of the distinction between initial-state (ISR) and 
final-state photon radiation (FSR). For ISR,  
$\MHV$ will be closer to the lepton-neutrino invariant mass and therefore
$\MHV^2 = (p_\ell + p_\nu)^2$ and 
$\pt = \pten$. For FSR, $\MHV$ will be closer to the invariant mass of the lepton-neutrino-gamma system: therefore, we choose 
$\MHV^2$ as $(p_\ell + p_\nu + p_\gamma)^2$ and the correct upper limit for partonic emission is here represented by the 
relative lepton-photon transverse momentum, for which we use the expression
\begin{equation}\label{eq:ptminlofsr}
p_\gamma \cdot p_\ell \, \frac{E_\gamma E_\ell}{(E_\gamma + E_\ell)^2}
\end{equation}
computed in the partonic CM system, for reasons that will be soon clarified.
 At the NLO accuracy, 
the $W\gamma$ inclusive cross section is treated according to the following formula~\cite{Hamilton:2012rf,Hamilton:2012np}
\begin{eqnarray}
\bar{B}_{W\gamma} &\longrightarrow &  \,  \Delta^2 \left( \MHV ,
  \pT \right) \left[ B_{W\gamma}\left( 1 - 2 \Delta^{(1)} \right) +\frac{\as(\pT)}{\as}
\left(V+ \int \frac{\mathd z}{z} G \right)\right]
\nonumber \\
 &+& \int \mathd \Phi_{\tmop{rad}} \frac{\as(\pT)}{\as} \Delta^2(\MHV,\pT)  R_{W\gamma;j}, 
\label{eq:sudagenwnlo}
\end{eqnarray}
where the meaning of $\MHV$ and $\pT$ is the same as in Eq.~(\ref{eq:sudagenw}), 
with the only difference that for the real radiation
matrix element $R \equiv |M_{W\gamma j}|^2 $ $\pT$ is calculated according to real radiation kinematics. 

Notice that the $\Delta^{(1)}$ term in eq.~(\ref{eq:sudagenwnlo}) contains Sudakov logarithms
that should cancel against similar logarithms arising in the last term of the equation
from the $\int \mathd \Phi_{\tmop{rad}}  R_{W\gamma;j}$ integral.\footnote{We
remind the reader that $R_{W\gamma;j}$ is singular, and that the singularities
are regulated by ``+'' distributions, so that by integrating it Sudakov logarithms can arise
as usual from phase space restrictions.}
 As far as FSR is concerned, in this
integral the radiation phase space is restricted
by the fact that  $R_{W\gamma;j}$ is suppressed when the radiation transverse momentum
is larger than the relative lepton-photon transverse momentum, defined according to the
default \POWHEGBOX{} internal mechanism for the separation of singular regions~\cite{Alioli:2010xd},
and corresponding precisely to the definition in eq.~(\ref{eq:ptminlofsr}), that thus
ensures that these large logarithms are fully cancelled, and correctly exponentiated.

Thanks to the MiNLO procedure, we are thus able to provide a 
generator that covers all the transverse momentum regions. We stress that, because of the 
non perturbative behaviour of the strong coupling constant in the low energy regime, a minimum cut on the 
parton or $W$ transverse momentum, say  $\pt^{j, {\rm min}} = 1$~GeV, must be 
necessarily introduced at the generation level. However, 
it does not play the r\^ole of a fictitious, unphysical cutoff as a consequence of the smooth and vanishing 
behaviour of the Sudakov form factor in the limit $\pt \to 0$, which renders the predictions independent 
of the specific $\pt^{j, {\rm min}}$ choice for sufficiently small $\pt^{j, {\rm min}}$ values.

\subsection{Interface to a shower generator}
The interface to the shower generator requires different kinds of vetoes in the
\POWHEG{}-C and in the \POWHEG{}-NC cases:
\begin{itemize}
\item \POWHEG{}-C: in this case \scalup{} is set to the hardest emission,
whether it is a parton (arising from the $W\gamma$ or $W j$  underlying Born) or
a photon ( $W j$ underlying Born). The mixed QED+QCD shower must be
turned on.
\item \POWHEG{}-NC: in this case, for events from the $W\gamma$ underlying
Born, \scalup{} is set to the hardest parton emission, as in the
default \POWHEG{} method. In the $W j$ underlying Born case, \scalup{}
should instead be set to the transverse momentum of the underlying Born
parton. It is not strictly necessary to turn on a full QED+QCD shower in this case,
since the hardest photon is already generated at the Les Houches event level.
If the QED shower is active, we must require that no photons are generated
harder then the Les Houches photon.
\end{itemize}

Shower MC generators do not in general enforce a veto on the transverse momentum
of photons radiated by the leptons, irrespective of the value of \scalup{}.
In \PYTHIA{}~v.8, it is possible to set a flag such that \scalup{} veto is
imposed also in this case.
In \PYTHIA{}~v.6 no such option exists.
We thus implement the photon veto at the analysis level.
We compute the relative transverse momentum of the lepton-photon system
after shower in the laboratory frame, defined as
\begin{equation}
p^{\rm rel}_{\ell\gamma}=2E_\gamma \sin\frac{\theta_{\ell\gamma}}{2}\,,
\end{equation}
for the first final-state shower generated photon arising from the lepton. 
 It is required that $p^{\rm rel}_{\ell\gamma}$ is smaller than \scalup{}, and the showering stage
is repeated keeping the same Les Houches event until this condition is satisfied.

\section{Phenomenological results}
\label{sec:pheno}
In the present Section, we show and discuss the numerical results obtained with the new tool. First, we show 
some comparisons with the NLO MC program \MCFM{}, which is the reference code for $W\gamma$ production
studies at the LHC. Then, we provide our phenomenological 
results for a number of differential cross sections of interest for physics studies at the Run~II of the LHC, 
and discuss the results of the different \POWHEG{}+MiNLO realizations
described in Section \ref{sec:he} and Section \ref{sec:minlo} in comparison to 
NLO calculations.
Finally, we compare our predictions with the measured cross sections at the LHC at 7 TeV.

\subsection{Comparisons to \MCFM{}: integrated cross sections}
\label{sec:mcfm}
We present a comparisons of our NLOPS generator and \MCFM{}
considering the following cases:
\begin{enumerate}
\item comparisons at pure NLO accuracy, using the smooth isolation approach as photon isolation procedure. 
This eliminates the fragmentation contribution and allows us to check our implementation of the NLO QCD 
corrections to direct photon production;
\item comparisons between the NLO predictions of \MCFM{} 
and the full results of our simulations.
In this comparisons we use a realistic photon isolation cut, and thus the \MCFM{}
result depends upon the photon fragmentation function.
We provide our predictions in terms of the \POWHEG{}-NC, \POWHEG{}-C-LO and \POWHEG{}-C-NLO
realizations.
\end{enumerate}
While in the first case we expect exact agreement, in the second case differences will show up
necessarily because of the different content of our approach.

The results presented here are obtained using the latest version of the \MCFM{} code, {\it i.e.} \MCFM{} v6.8. 
We used the following set of EW input parameters
\begin{eqnarray}
&& M_W = 80.385~{\rm GeV} \quad M_Z = 91.1876~{\rm GeV} \,\,\,\, \, \, G_\mu = 1.1663787 \times 10^{-5}~{\rm GeV}^{-2}  \\
&& \Gamma_W = 2.085~{\rm GeV} \quad \quad \alphaem (0) = 1/137.035999074
\end{eqnarray}
We compute the $O(\alphaem^3)$ LO cross section using for the electromagnetic coupling constant the expression
\begin{eqnarray}
\alpha_{G_\mu} \, = \, \frac{\sqrt{2} \, G_\mu \, M_W^2 \sin^2\theta_W}{\pi} \; ,
\end{eqnarray}
with $\sin^2\theta_W = 1 - M_W^2/M_Z^2$, and we rescale the results by the factor 
$\alphaem (0)/\alpha_{G_\mu}$ to account for the correct coupling to the on-shell photon.
For definiteness, we use the NLO PDF set CTEQ10~\cite{Lai:2010vv}
(any other modern set can equivalently be
used~\cite{Martin:2009iq,Ball:2010de}).
For the fragmentation of partons into
photons, needed by the \MCFM{} code,
the parameterization ``Set II" of ref.~\cite{Bourhis:1997yu} was used.

Similarly to the analysis performed in ref.~\cite{Campbell:2011bn}, as well as to 
cover the main event selection conditions of interest at the LHC, we consider 
both in comparison 1 and 2 the following three sets of cuts
\begin{subequations}\label{eq:cuts}
\begin{align}
\label{eq:cut1}
{\rm Basic \, \, Photon} & - \, \, \pt^\gamma > 15~{\rm GeV}, \, |\eta_\gamma| < 2.37,  \, \Delta R_{\ell\gamma} > 0.7, \, R_0 = 0.4, \, \epsilon_h = 0.5 \\
\label{eq:cut2}
M_T {\rm \, \, cut} & - \, \, {\rm Basic \, \, Photon} + M_T > 90~{\rm GeV}\\
\label{eq:cut3}
{\rm Lepton \, \, cuts} & - \, \, {\rm Basic \, \, Photon}  + \pt^\ell > 25~{\rm GeV}, \,  |\eta_\ell| < 2.47, \, \pt^\nu > 35~{\rm GeV}\;,
\end{align}
\end{subequations}
where $R_0$ and $\epsilon_h$ are the parameters defining the isolation criterion, as defined further on.
They give rise to quite different $K$-factors~\cite{Campbell:2011bn} and therefore allow us to perform a non-trivial test of our calculation. 
In Eq.~(\ref{eq:cut2}), $M_T$ is the transverse mass of the photon-lepton-missing transverse energy system,
 defined as
 \begin{equation}
 M_{T, \ell\nu\gamma}^2 \, = \, \left(\sqrt{m_{\ell\gamma}^2 + |  \bold{\pt^{\gamma}} + \bold{\pt^{\ell}} |^2 } + E_T^\nu \right)^2 - 
 | \bold{\pt^{\gamma}} +  \bold{\pt^{\ell}} + \bold{E_T^{\nu}} |^2 \; ,
 \end{equation}
where $m_{\ell\gamma}$ is the invariant mass of the photon-lepton system. A $M_T$ cut is particularly useful 
since it suppresses the contribution of photons radiated by the lepton in the $W$ decay, which is of no 
interest for the studies of ATGCs and for the observation of radiation zeros. The basic photon cuts defined
in Eq.~(\ref{eq:cut1}) mimic the criteria adopted by the 
CMS collaboration in the comparison of \MCFM{} to the data at $\sqrt{s}$ = 7~TeV,\footnote{Strictly speaking, 
in the criteria adopted by CMS no acceptance cuts on the lepton are applied and the photon 
isolation requirements are more complex than those as in Eq.~(\ref{eq:cut1}). } while the lepton
in Eq.~(\ref{eq:cut3}) resemble the experimental configuration used by ATLAS 
in the comparison between data and theoretical predictions at $\sqrt{s}$ = 7~TeV. Moreover, the 
conditions in Eq.~(\ref{eq:cut3}), including an additional cut on the transverse mass of the 
$\ell \nu \gamma$ or $\ell \nu$ system, are similar to the selection criteria applied by ATLAS and CMS, respectively, 
to define the $W\gamma$ sample.
\subsubsection{NLO comparison}
We present results for the final state 
$e^+ \nu \gamma$, using a central scale choice of $\mu_R = \mu_F = M_W$. 
The fragmentation scale is kept equal to $M_W$.
We apply the following isolation 
prescription to the photon~\cite{Frixione:1998jh}
\begin{equation}
\forall_{R<R_0} \sum_{R_{j,\gamma} < R} \, E_{T,j} < \epsilon_h \, \pt^\gamma \left( \frac{1 - \cos R}{1 - \cos R_0} \right) \; ,
\end{equation}
where $E_{T,j}$ is the transverse energy of the final state parton $j$, and
$R_{j\gamma} = \sqrt{\Delta\eta_{j\gamma}^2 + \Delta\phi_{j\gamma}^2}$
is the ``separation" between the photon and a parton $j$.
In case of the NLO calculation, only one parton is present in the final state, and the
isolation condition can be simply stated as
\begin{equation}
\theta(R_0-R_{j,\gamma})\, E_{T,j}< \epsilon_h \, \pt^\gamma \left( \frac{1 - \cos R_{j,\gamma}}{1 - \cos R_0} \right) \; ,
\end{equation}
where $j$ here stands for the single final state parton.
 We used the values $\epsilon_h = 0.5$ and $R_0 = 0.4$
for the isolation parameters.

The results of the comparisons
are shown in Tab.~\ref{tab:NLO7} and~\ref{tab:NLO14} for $\sqrt{s} = 7$ TeV and $\sqrt{s} = 14$ TeV, 
respectively. As can be seen, the predictions of \MCFM{} and \POWHEG{} agree within the statistical uncertainty of the 
respective MC errors, given by the numbers in parenthesis. For completeness, we also checked that 
the distributions obtained with the two calculations are in good agreement. Thus, in the absence of the fragmentation 
contribution, the two calculations of the NLO QCD corrections to $W\gamma$ production nicely agree.
\begin{table}[t]
\centering
\begin{tabular}{| c || c | c |}
\hline
Cuts & \MCFM{} & \POWHEG{} NLO \\
\hline
Basic Photon & 13.12(4) & 13.15(1) \\
\hline
$M_T$ cut & 2.770(1) & 2.774(3) \\
\hline
Lepton cuts & 1.126(1) & 1.123(4) \\
\hline
\end{tabular}
\caption{\label{tab:NLO7}  Comparison between \MCFM{} and \POWHEG{} NLO cross sections (in pb) 
of the $p p \to e^+ \nu \gamma$ process at $\sqrt{s}$ = 7~TeV, 
using the smooth isolation procedure, parameters and cuts described in the text. The numbers in parenthesis are the 
1$\sigma$ MC errors on the last digit. The results are the central value predictions for 
$\mu_R = \mu_F = M_W$.}
\end{table}
\begin{table}[t]
\centering
\begin{tabular}{| c || c | c |}
\hline
Cuts & \MCFM{} & \POWHEG{} NLO \\
\hline
Basic Photon & 23.90(8) & 24.04(3) \\
\hline
$M_T$ cut & 6.230(2) & 6.250(9) \\
\hline
Lepton cuts & 2.342(2) & 2.340(6) \\
\hline
\end{tabular}
\caption{\label{tab:NLO14}  The same as Tab.~\ref{tab:NLO7} at $\sqrt{s}$= 14 TeV.}
\end{table}

\subsubsection{Full comparison}
In our full comparison between our NLOPS simulations and the \MCFM{} results,
photon isolation is implemented by requiring that the transverse hadronic energy inside $R_0$ is limited by the 
following condition
\begin{equation}
\sum_{R_{j,\gamma} < R_0} \, E_{T,j} < \epsilon_h \, \pt^\gamma \; ,
\end{equation}
where $j$ runs over all the final state particles except the photon and the electron,
with $R_0 = 0.4$ and $\epsilon_h = 0.5$, as before. 
The shower program ensures that no QCD radiation
will be generated harder than the \scalup{} value. We enforce an analogous limitation
on the radiation of photons: for a given Les Houches event, we repeat the shower
stage if a photon harder then \scalup{} is generated.
We do not take into account in our simulations the contribution of secondary photons from the
main radiative decays of hadrons, because they are normally treated as background.

In order to make contact with the \MCFM{} formulation, 
we switch off in the shower generator the contribution of the underlying event, 
thus including only the shower and hadronization stages. In order to further minimize the possible sources 
of discrepancies, we require in our simulations that the charged lepton defining the signature
is identified as the hardest lepton in the event sample and assume that the isolated photon 
coincides with the hardest photon among all the isolated ones\footnote{We checked in our simulations 
that no appreciable differences are present if the lepton, as well as the isolated photon (whenever possible), 
are identified according to the MC truth.}. Moreover, we
require a calorimetric-like definition for the final state lepton, {\it i.e.} a lepton+photon 
recombination requirement for the photons generated by the QED PS, 
consistently with the dressed lepton definition inherent in \MCFM{}. The latter requirement 
is imposed by setting the parameter {\tt PTminChgL} = 1~GeV in \PYTHIA{}~v.8. Our results are obtained with
\PYTHIA{} v.8 as PS generator but we checked that no substantial differences (at $\sim 1\%$ level) are introduced when using 
\PYTHIA{} v.6.

The comparisons between our full predictions and those of \MCFM{}
are given in Tab.~\ref{tab:Full7} and Tab. \ref{tab:Full14}
for $\sqrt{s} = 7$ TeV and $\sqrt{s} = 14$ TeV, respectively. We include the \MCFM{} theoretical uncertainties 
arising from the scale dependence. We set the factorization and renormalization scales
equal to  $\mu_F = K_F \, \mu_0$ and $\mu_R = K_R \, \mu_0$, 
with $\mu_0 = M_W$,\footnote{We checked that no substantially different conclusions 
derive from the choice of a dynamical factorization/renormalization scale as central value in \MCFM{}.}
and evaluate the cross sections for the following choices
\begin{eqnarray}
\!\!\!\! (K_F, K_R) \!  \in  \! \left\{ \left( \frac{1}{2} , \frac{1}{2} \right), \left( \frac{1}{2} , 1 \right), \left(\frac{1}{2},2 \right), \left( 1, \frac{1}{2} \right), \left( 1, 1 \right),
\left( 1, 2 \right), \left( 2, \frac{1}{2} \right),  \left( 2, 1 \right),  \left( 2, 2 \right)
\right\} \label{eq:scales}
\end{eqnarray}
that include variations of the two scales in the same and opposite directions.
In fact,  as  motivated in  ref.~\cite{Campbell:2011bn}, 
varying the scales in opposite directions leads to a more 
reliable estimate of the theoretical uncertainty.
The upper and lower values of the cross section that we quote are the upper and lower limits of the
scale variation. 
The fragmentation scale is kept equal to $M_W$, as its variation does not lead to a significant change in the
results, as discussed in ref.~\cite{Campbell:2011bn}.
Concerning our results, which rely upon a dynamical 
treatment of the scales according to the MiNLO procedure, as described in Section \ref{sec:minlo}, 
the factorization scale variation is introduced both in the $W\gamma$ NLO and $Wj$ LO/NLO 
dynamics (see Eqs.~(\ref{eq:sudagen}), (\ref{eq:sudagenwjnlo}), (\ref{eq:sudagenw}) and (\ref{eq:sudagenwnlo})) 
according to the procedures described in refs.~\cite{Hamilton:2012rf,Hamilton:2012np},
where a factor $K_R$ is introduced that multiplies the
MiNLO nodal scales, and a factor $K_F$ is associated with the factorization scale.

In Tab.~\ref{tab:Full7} and Tab. \ref{tab:Full14} the upper and lower extrema  are obtained by calculating 
the cross section for both \MCFM{} and \POWHEG{}+MiNLO at $\{ K_R =1/2 , K_F = 2 \}$ and $\{ K_R = 2 , K_F = 1 / 2 \}$, respectively,
in agreement with the conclusion of ref.~\cite{Campbell:2011bn}. 

\begin{table}[t]
\centering
\begin{tabular}{| c || c | c | c | c|}
\hline
Cuts & \MCFM{} & \POWHEG{}-NC & \POWHEG{}-C-LO & \POWHEG{}-C-NLO\\
\hline
Basic Photon & 12.92(3)$^{+4\%}_{-6\%}$ & 12.40(3)$^{+8\%}_{-10\%}$ &  12.95(3)$^{+8\%}_{-11\%}$ & 15.08(7)$^{+2\%}_{-9\%}$ \\
\hline
$M_T$ cut & 2.625(1)$^{+6\%}_{-6\%}$  & 3.09(2)$^{+10\%}_{-11\%}$ &  3.20(2)$^{+11\%}_{-11\%}$ & 4.23(3)$^{+10\%}_{-10\%}$ \\
\hline
Lepton cuts & 1.077(1)$^{+6\%}_{-6\%}$ & 1.22(1)$^{+8\%}_{-10\%}$ & 1.31(1)$^{+11\%}_{-11\%}$ & 1.75(2)$^{+7\%}_{-13\%}$ \\
\hline
\end{tabular}
\caption{\label{tab:Full7} 
Comparison between the NLO cross section predictions (in pb) of \MCFM{} using photon fragmentation functions 
and the three different \POWHEG{}+MiNLO implementations realized in this work, for the 
$p p \to e^+ \nu \gamma$ process at $\sqrt{s}$ = 7~TeV. Photon isolation 
conditions, parameters and cuts are specified in the text. The basic theoretical ingredients 
underlying the acronyms \POWHEG{}-NC, \POWHEG{}-C-LO and \POWHEG{}-C-NLO are summarized 
in the present Section. The numbers in parenthesis are the statistical errors on the last digit. 
The uncertainties are estimated from the scale dependence, as explained 
in the text. }
\end{table}

\begin{table}[t]
\centering
\begin{tabular}{| c || c | c | c | c|}
\hline
Cuts & \MCFM{} & \POWHEG{}-NC & \POWHEG{}-C-LO & \POWHEG{}-C-NLO\\
\hline
Basic Photon & 23.47(1)$^{+5\%}_{-8\%}$ & 22.57(7)$^{+10\%}_{-15\%}$ & 23.52(7)$^{+11\%}_{-16\%}$ & 28.51(13)$^{+4\%}_{-11\%}$  \\
\hline
$M_T$ cut & 5.839(1)$^{+9\%}_{-9\%}$ & 6.99(4)$^{+14\%}_{-15\%}$ & 7.11(4)$^{+15\%}_{-16\%}$ & 9.99(8)$^{+14\%}_{-14\%}$ \\
\hline
Lepton cuts & 2.227(1)$^{+9\%}_{-10\%}$ & 2.55(2)$^{+14\%}_{-15\%}$  &  2.67(2)$^{+16\%}_{-17\%}$ & 3.76(5)$^{+17\%}_{-13\%}$ \\
\hline
\end{tabular}
\caption{\label{tab:Full14} 
The same as Tab. \ref{tab:Full7} at $\sqrt{s}$ = 14~TeV}
\end{table}

Before discussing the main aspects of these comparisons, let us remind, for clarity, the basic theoretical ingredients underlying the
three \POWHEG{}+MiNLO implementations realized in our work and considered in  Tab. \ref{tab:Full7} and Tab. \ref{tab:Full14}.
As detailed in Section \ref{sec:he}, the three variants are characterized by the following main features:

\begin{itemize}

\item \POWHEG{}-NC
\begin{enumerate}
\item Accuracy: $\bar{B} = \bar{B}_{W\gamma} + {B'}_{Wj}$, with NLO QCD accuracy for $W\gamma$ production and 
LO accuracy for the $Wj$ subprocess.
\item Radiation dynamics: QCD radiation $R_{W\gamma;j}$ in $W\gamma$ production and QED 
radiation $R_{Wj;\gamma}$ in $Wj$ contribution.
\item \scalup{}: as in the default \POWHEG{} method for the $\bar{B}_{W\gamma}$ events and
equal to the transverse momentum of the parton for the $B_{Wj}$ contribution.
\end{enumerate}
\item \POWHEG{}-C-LO
\begin{enumerate}
\item $\bar{B} = \bar{B}_{W\gamma} + {B'}_{Wj}$, like for \POWHEG{}-NC.
\item Radiation dynamics: QCD radiation in $W\gamma$ production and
QED+QCD emission in the $Wj$ contribution.
\item \scalup{}: as in the default \POWHEG{}, for both QCD radiation in the $\bar{B}_{W\gamma}$ events,
and QED and QCD emission in the ${B'}_{Wj}$ ones.
\end{enumerate}
\item \POWHEG{}-C-NLO
\begin{enumerate}
\item $\bar{B} = \bar{B}_{W\gamma} + \bar{B}_{Wj}$, with NLO QCD accuracy for both $W\gamma$ and 
$Wj$ contributions.
\item Radiation dynamics: the same as in \POWHEG{}-C-LO.
\item \scalup{}: the same as in \POWHEG{}-C-LO.
\end{enumerate}
\end{itemize}
All the above ingredients are supplemented with the MiNLO procedure detailed in Section~\ref{sec:minlo} 
and by an interleaved QCD+QED simulation of multiple QCD and photon radiation.

As can be noticed from the results shown in Tab. \ref{tab:Full7} and Tab. \ref{tab:Full14}, different considerations
can be made, depending to a large extent on the assumed experimental setup. 

First, it is evident that the theoretical uncertainties are pretty large, in particular at $\sqrt{s}$ = 14~TeV, 
and this is a first hint of the importance of higher-order corrections to $W\gamma$ production at the LHC. 
In particular, it can be seen that the uncertainties associated to the \POWHEG{}+MiNLO predictions 
are larger than those of \MCFM{}, as a consequence of the quite different procedure used 
in the estimate of the theoretical uncertainty from the scale variations. In particular, 
our results receive an additional uncertainty driven by the evaluation of the $Wj$ LO matrix elements 
and the $Wj(j)$ NLO cross section at relatively small transverse momenta.

In the presence of basic photon cuts only, the predictions of \POWHEG{}-C-LO are in fairly good 
agreement with those of \MCFM{} at both LHC energies. The cross sections obtained with the
\POWHEG{}-NC method are slightly lower, at the level of some percent, than those obtained 
with \POWHEG{}-C-LO. The same behavior is observed in the $M_T$ or lepton cut selections, 
leading to the conclusion that the \POWHEG{}-NC method is a good approximation of the 
\POWHEG{}-C-LO one. On the other hand, the full NLO corrections to the $Wj$ contribution, included 
in the \POWHEG{}-C-NLO predictions, increases the cross section central values obtained with 
\POWHEG{}-C-LO by about 15--20\% in the basic photon selection, due to the 
presence of higher-order (beyond $O(\as)$) QCD contributions in our most accurate calculation. 

In the presence of additional $M_T$ or lepton cuts, different conclusions can be drawn. For such 
configurations, it is known that the NLO $K$-factors to the $W\gamma$ cross section at the LHC are 
quite large~\cite{Campbell:2011bn}, approaching a factor of two. 
Under these conditions, the results obtained in terms of all our realizations are 
substantially different from the \MCFM{} predictions. These discrepancies can be 
attributed to a number of reasons: the presence of PS effects and the 
modeling of the fragmentation contribution in all our algorithms, as well as the 
inclusion of  the NLO corrections to the $Wj$ subprocess in our
\POWHEG{}-C-NLO calculation.

The presence of large NLO corrections to the $Wj$ process may be viewed
as an indication that higher-order QCD contributions and, in particular, 
the yet unavailable NNLO corrections\footnote{Preliminary results of a complete calculation 
of QCD corrections to the $W\gamma$ process at NNLO accuracy have been presented in ref.~\cite{Grazzini:2014pqa}. 
For a setup close to ATLAS 
analysis at 7 TeV, but using a smooth cone isolation requirement, the NNLO corrections amount to
about +20\%.} 
may play a prominent r\^ole for a reliable theoretical interpretation of $W\gamma$ data at the LHC. 
On the other hand, we remind the reader that the $Wj$ generated events may be view as an independent
part of the production phase space, corresponding to the case when the emitted photon is softer than
at least one jet in the event, while for $W\gamma$ generated events the photon is harder than all
jets. Thus, it makes sense to include the NLO corrections to the $Wj$ generated events, even if we do not attempt to include
higher order corrections to the $W\gamma$ ones. Since these corrections are applied in different phase
space regions, there is no reason to expect cancellations among them.
\begin{table}[t]
\centering
\begin{tabular}{| c || c | c |}
\hline
Photon isolation & \POWHEG{}-C-LO [$\sqrt{s} =7$~TeV] & \POWHEG{}-C-LO [$\sqrt{s} =14$~TeV] \\
\hline
$R_0 = 0.4$, $\epsilon_h = 0.4$ & 1.30(1) & 2.64(2) \\
\hline
$R_0 = 0.4$, $\epsilon_h = 0.5$ & 1.31(1) & 2.67(2) \\
\hline
$R_0 = 0.4$, $\epsilon_h = 0.6$ & 1.32(1) & 2.70(2) \\
\hline\hline
$\epsilon_h = 0.5$, $R_0 = 0.3$  & 1.34(1) & 2.75(2) \\
\hline
$\epsilon_h = 0.5$, $R_0 = 0.5$ & 1.29(1) & 2.60(2) \\
\hline
\end{tabular}
\caption{\label{tab:isolation} Variation of the \POWHEG{}+MiNLO predictions (\POWHEG{}-C-LO implementation
 under lepton cuts conditions) for 
the cross section (in pb) of the $p p \to e^+ \nu \gamma$ process at $\sqrt{s}$ = 7~TeV and $\sqrt{s}$ = 14~TeV, respectively, 
due to a change of the photon isolation parameters $\epsilon_h$ and $R_0$.}
\end{table}

Regardless of the comparison to \MCFM{}, it is also worth noting that the predictions provided 
by the algorithm \POWHEG{}-NC and \POWHEG{}-C-LO 
are in agreement at the some percent level both in the basic photon conditions and 
in the more exclusive $M_T$ and lepton cut selections. While in the first algorithm,  which mimics the approach 
of ref.~\cite{D'Errico:2011sd}, the QCD/QED interplay is modeled 
using an appropriate \scalup{} choice for the QCD shower, in the second algorithm the
radiation dynamics is described in a manner fully consistent with the default \POWHEG{} method, 
using the correct Sudakov form factors for both photon and parton emissions from an underlying
$Wj$ Born configuration. The improved dynamical description of the radiation mechanism 
leads to some appreciable differences, even if we can conclude that the method proposed 
by D'Errico and Richardson in ref.~\cite{ D'Errico:2011sd} is already a realistic approximation 
of the our \POWHEG{}-C-LO implementation.

To conclude this Section, we show in Tab.~\ref{tab:isolation} the variation of a sample of our 
cross section results obtained by changing the parameters
that define the photon isolation criteria. We restrict ourselves to the predictions of the 
\POWHEG{}-C-LO implementation under lepton 
cuts conditions, as we checked that very similar conclusions hold for the other realizations and experimental setup. 
We study the variation induced by a mild change of both the fraction $\epsilon_h$ of the 
transverse energy inside the cone and the cone size $R_0$.
As can be seen, our results are stable against variations of $\epsilon_h$ and $R_0$ both at $\sqrt{s}$ = 7~TeV 
and $\sqrt{s}$ = 14~TeV, as the cross section changes are of the order of a few percent. This suggests 
that the fragmentation contribution has a rather moderate impact on total prompt-photon production, 
in the presence of realistic photon isolation criteria.

\subsection{Differential cross sections  at $\sqrt{s}$ = 14~TeV:  NLO vs. NLOPS predictions}
\label{sec:diff}
In the studies of the $W\gamma$ process at the LHC, several distributions are considered
in order to test SM predictions as well as to 
set limit on ATGCs and new vector resonance production \cite{Aad:2013izg,Chatrchyan:2013fya}.
Here we limit ourselves to present our predictions for a particularly 
representative sample of differential cross sections at $\sqrt{s}$ = 14~TeV. Moreover, we focus
on the results for the basic photon and lepton cuts only, as the considerations valid for 
the $M_T$ cut conditions are similar to the latter case.

We show and comment the results obtained with all the three variants of our \POWHEG{} implementations
for the following distributions:
\begin{itemize}

\item the photon and lepton $\pt$;

\item the transverse mass of the three-body system $e^+\nu\gamma$, {\it i.e.} $M_T^{\ell\nu\gamma}$;

\item the photon-lepton rapidity difference $\Delta\eta (e^+\gamma)$;

\end{itemize}
and we compare our NLOPS simulations to the NLO \MCFM{} predictions, including the \MCFM{} theoretical uncertainties 
obtained from the scale dependence, estimated according to the method discussed for the total cross section
(see eq.~\ref{eq:scales}).\footnote{As for the comparisons at the level of 
integrated cross sections, we checked that the choice of a dynamical factorization/renormalization
scale as central value in \MCFM{} does not lead to substantially different conclusions.}
The uncertainty bands that we quote are the envelope of the results obtained with the
different scale choices.
\begin{figure}[t]
\begin{center}
\includegraphics[width=7.5cm]{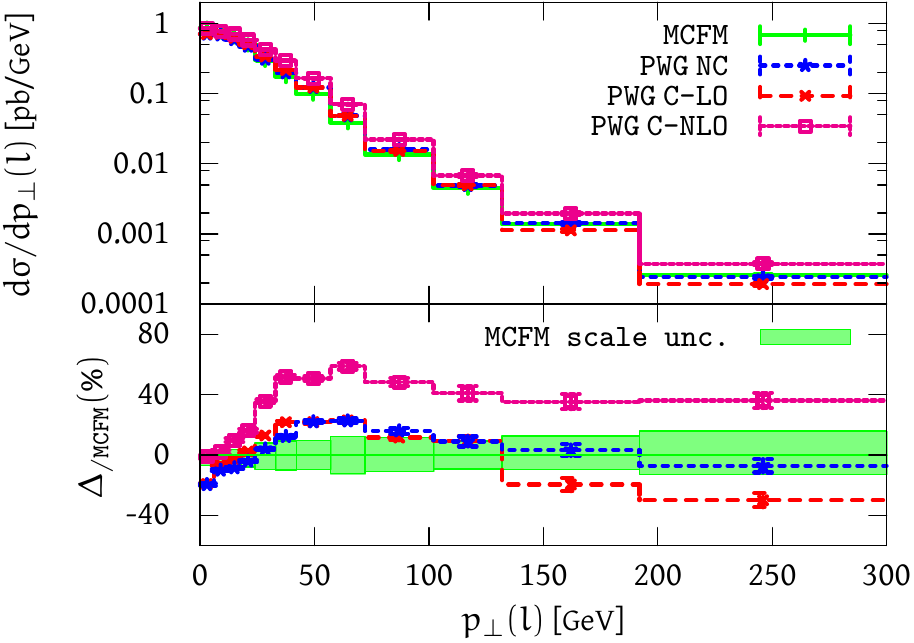}~~~~~\includegraphics[width=7.5cm]{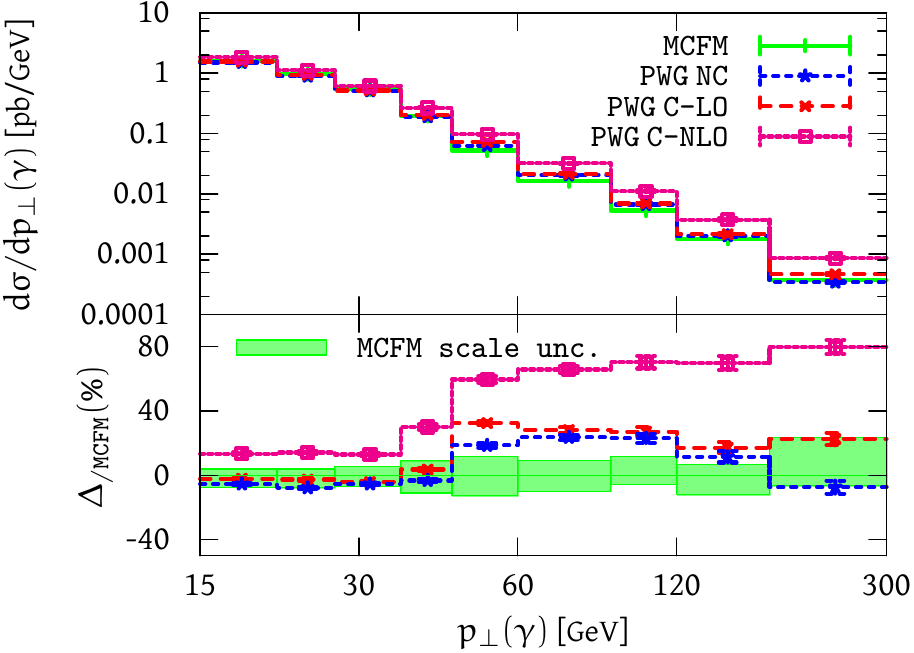}
\includegraphics[width=7.5cm]{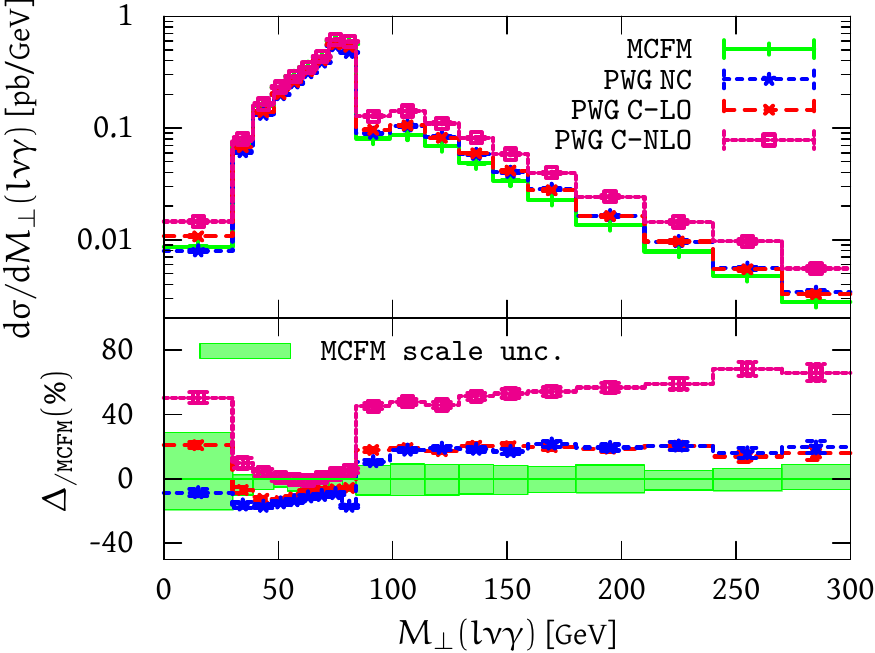}~~~~~~~~~\includegraphics[width=7.15cm]{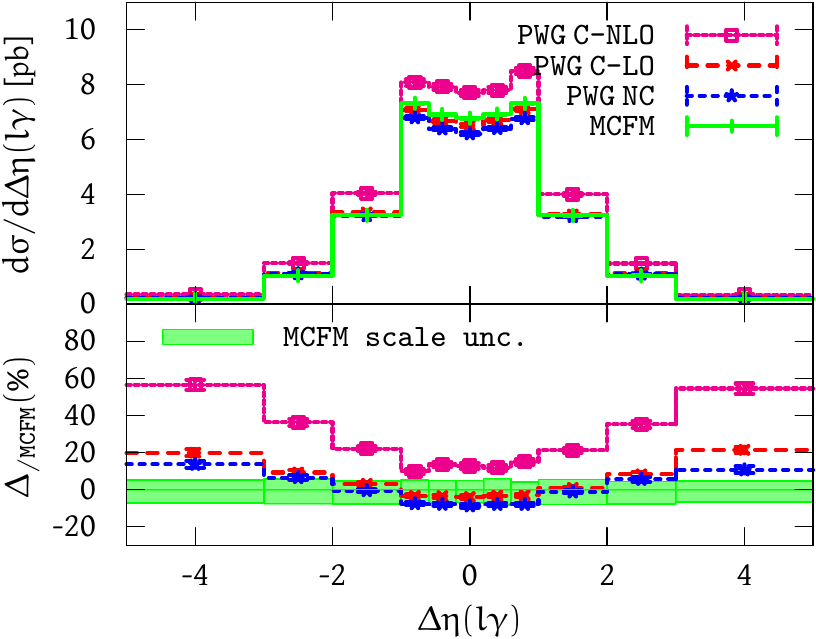}
\end{center}
\caption{\label{Basic} Comparison between the NLOPS results of the three \POWHEG{}+MiNLO realizations and the 
\MCFM{} NLO predictions for a number of distributions under basic photon cuts conditions. Upper plots: $\pt^\ell$ and 
$\pt^\gamma$ distributions; lower plots: $M_T^{\ell \nu \gamma}$ and $\Delta\eta (\ell\gamma)$ distributions. 
The lower panels show $\Delta / \MCFM = (d\sigma_{\PWG} - d\sigma_{\MCFM}) /d\sigma_{\MCFM}$, where the 
green band is the \MCFM{} theoretical uncertainty obtained from the 
scale variation, as explained in the text. }
\end{figure}
\begin{figure}[h]
\begin{center}
\includegraphics[width=7.5cm]{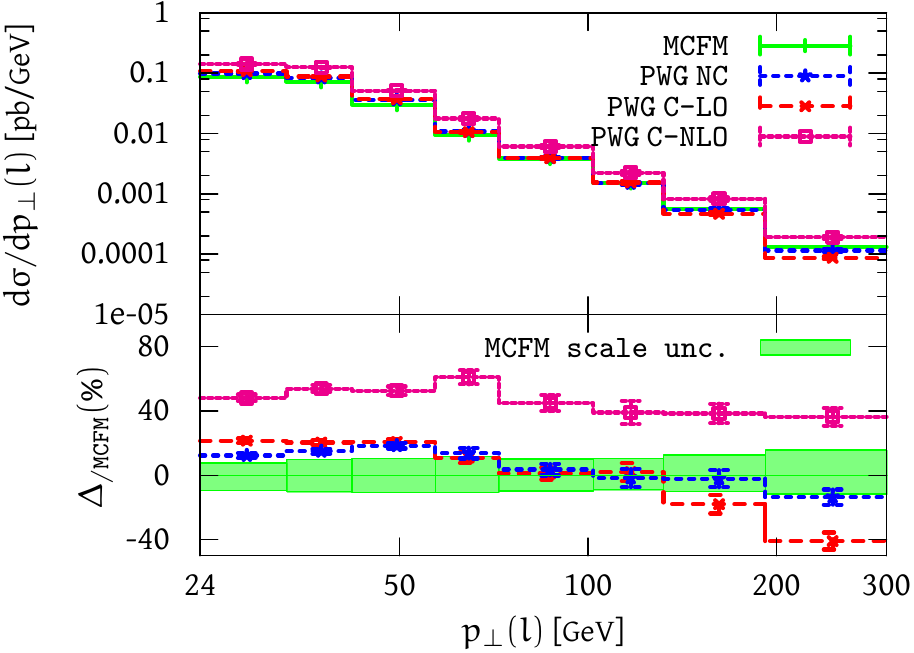}~~~~~\includegraphics[width=7.5cm]{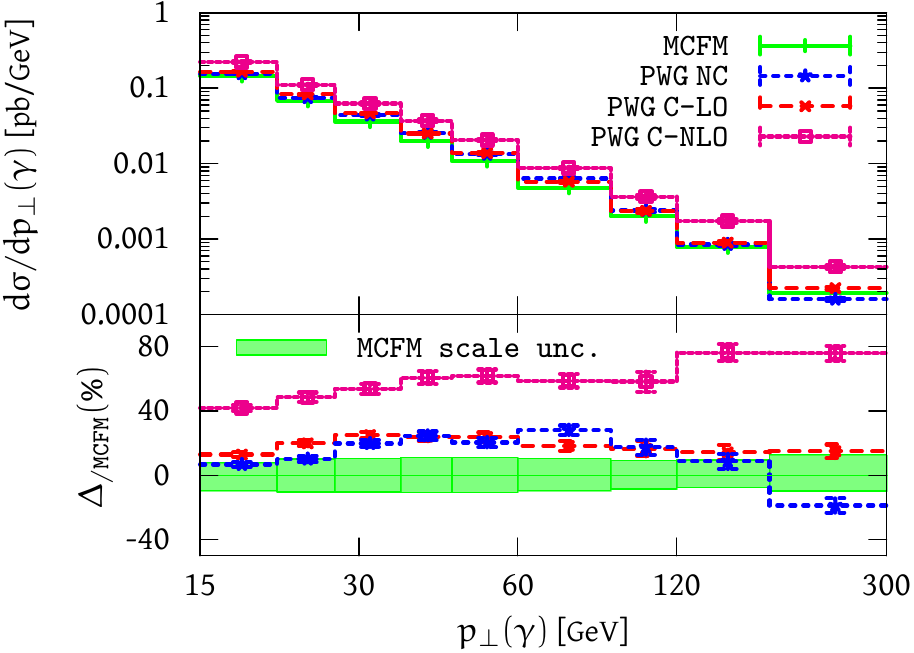}
\includegraphics[width=7.5cm]{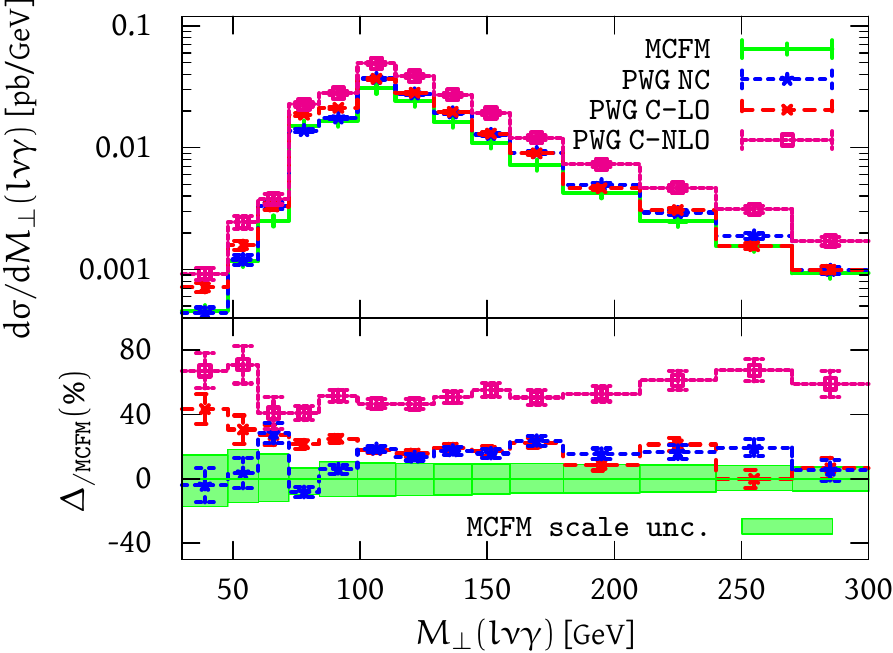}~~~~~~~~~\includegraphics[width=7.15cm]{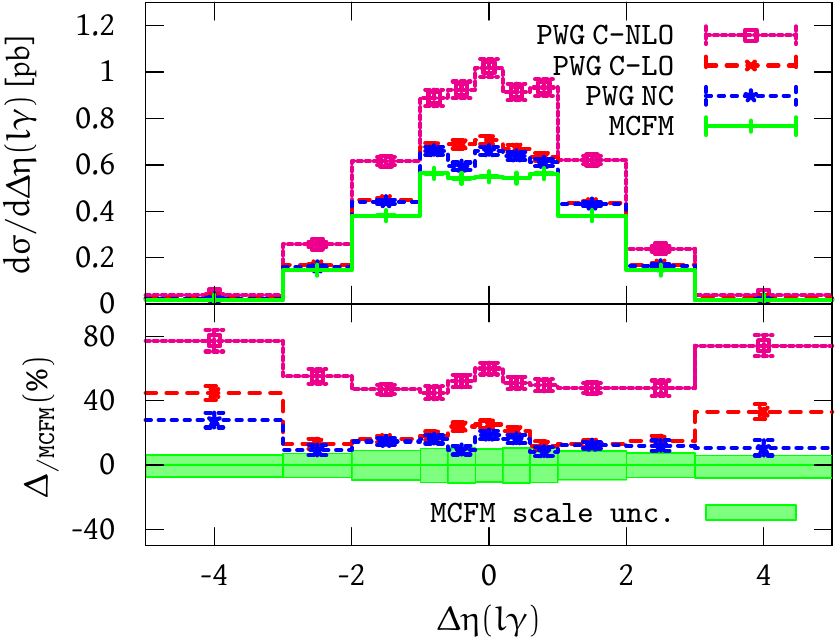}
\end{center}
\caption{\label{Lepton} The same as Fig. \ref{Basic} under lepton cuts conditions.}
\end{figure}
The results of these comparisons are shown in Fig. \ref{Basic} and Fig. \ref{Lepton} for the basic photon and lepton cuts 
conditions, respectively. The \MCFM{} uncertainties are shown in the lower panel of each plot, where the
quantity $\Delta / \MCFM$ is defined as $\Delta / \MCFM =  (d\sigma_{\PWG} - d\sigma_{\MCFM}) /d\sigma_{\MCFM}$, 
thus providing the relative deviation between our results and the NLO \MCFM{} predictions.


For the basic photon setup, the shape of the differential cross sections predicted by MCFM and our NLOPS algorithms 
are rather different, in spite of the good agreement observed between MCFM
and our results given by POWHEG-NC and POWHEG-C-LO at the level of integrated cross sections.
This can be ascribed to the several different elements that are present in our calculation and are not
included in the fixed order calculation. In particular, the region of small lepton momenta displays
a marked difference in shape, since it is affected by our treatment of the separation between
$W\gamma$ and $W j$ underlying Born processes, with the consequent use of the MiNLO procedure.
On the other hand, it is apparent that  the largest differences
are observed with respect to the \POWHEG{}-C-NLO scheme, due to the fact that in this scheme we include
large QCD higher order corrections to the $W j$ process that are not at all present in the fixed order
calculation.

From the lower panels of the
plots shown in Fig. \ref{Basic} it can be also seen that 
some large variations show up in the relative deviations of our predictions to those of \MCFM{}. They are consequence of 
crossing points present in the shape of the distributions obtained in the two approaches.
For example, in the $p_\perp^\gamma$ spectrum, the raise of the \POWHEG{} result with respect to the
\MCFM{} one around $50-60$ GeV can be ascribed to the fact that photons from $W$ decays are less likely
above that value, and the two approaches have a very different treatment of the fragmentation contribution.
Therefore, the main message of this comparison is that NLO and NLOPS
calculations predict substantially different shapes for the differential cross sections of the $p p \to \ell \nu \gamma$ 
process when inclusive experimental conditions, {\it i.e.} without cuts on the lepton variables or on the transverse invariant
masses, are considered. 

A further conclusion that can be drawn from inspection of Fig. \ref{Basic} is that the simulations 
obtained with the two algorithms \POWHEG{}-NC and \POWHEG{}-C-LO are in good agreement 
in the dominant regions of the distributions, thus explaining the agreement already noticed 
at the level of the integrated cross sections. 
On the other hand, the inclusion of the $Wj$ contribution with normalization at NLO accuracy as in \POWHEG{}-C-NLO gives 
rise to a relevant effect on the normalization of the distributions, outside the \MCFM{} theoretical uncertainty band. 
The latter conclusion also
holds for the distributions shown in Fig.~\ref{Lepton} under lepton cuts conditions. However, in this
case one can notice that the ratio of our predictions to those of \MCFM{} is more uniform,
as we have eliminated the differences that arise in the region of small lepton transverse momenta.
 Nevertheless, it can be seen that the predictions of 
all our \POWHEG{} implementations are substantially different from those of \MCFM{} at the normalization level. 
\begin{figure}[t]
\begin{center}
\includegraphics[width=7.5cm]{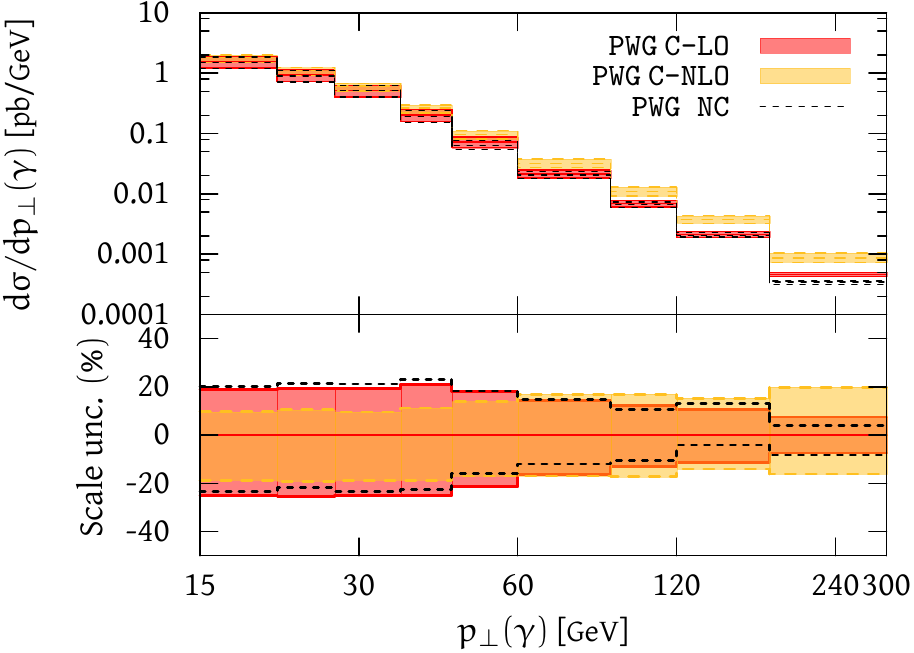}~~~~~\includegraphics[width=7.5cm]{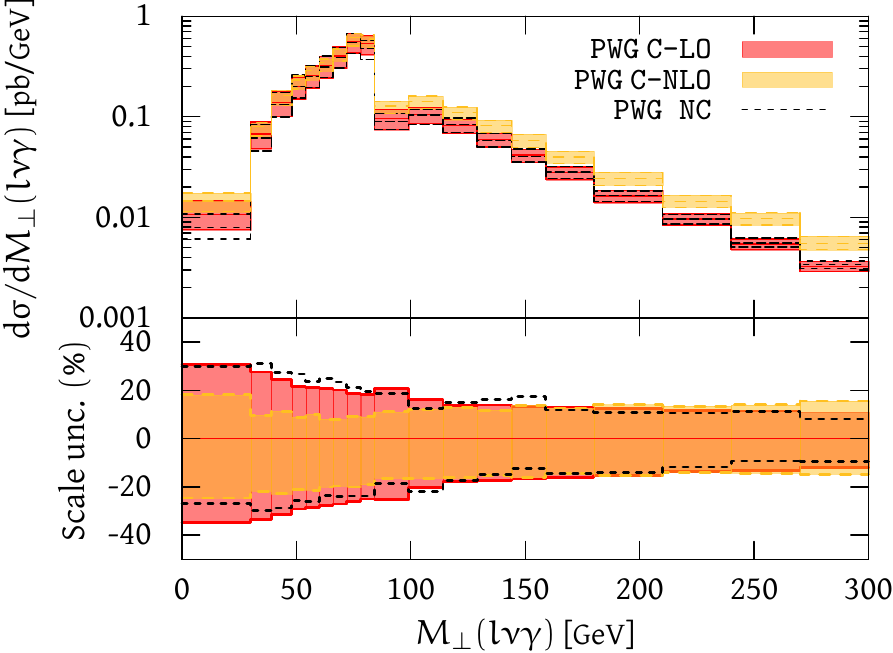}
\includegraphics[width=7.5cm]{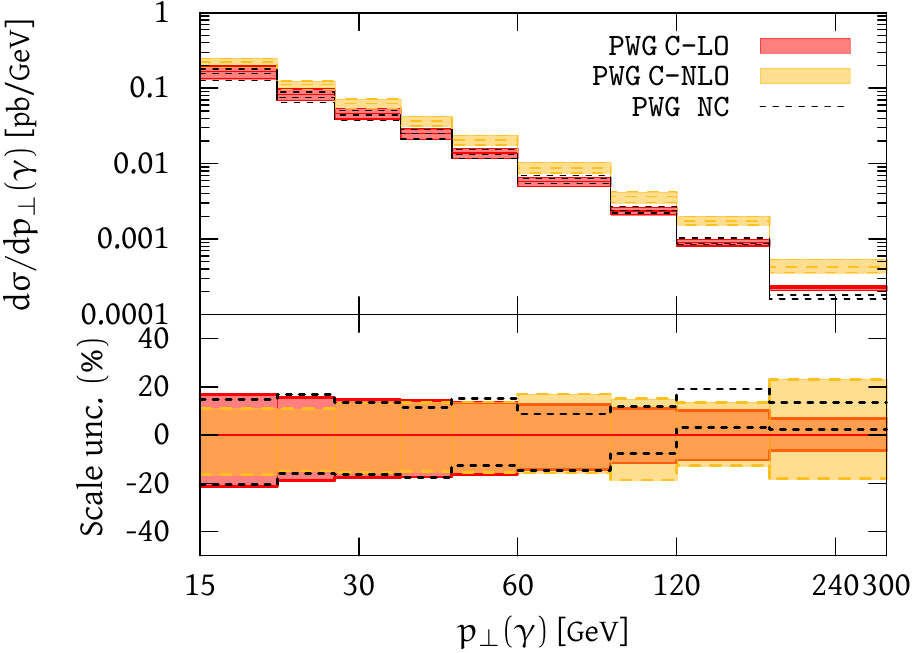}~~~~~\includegraphics[width=7.5cm]{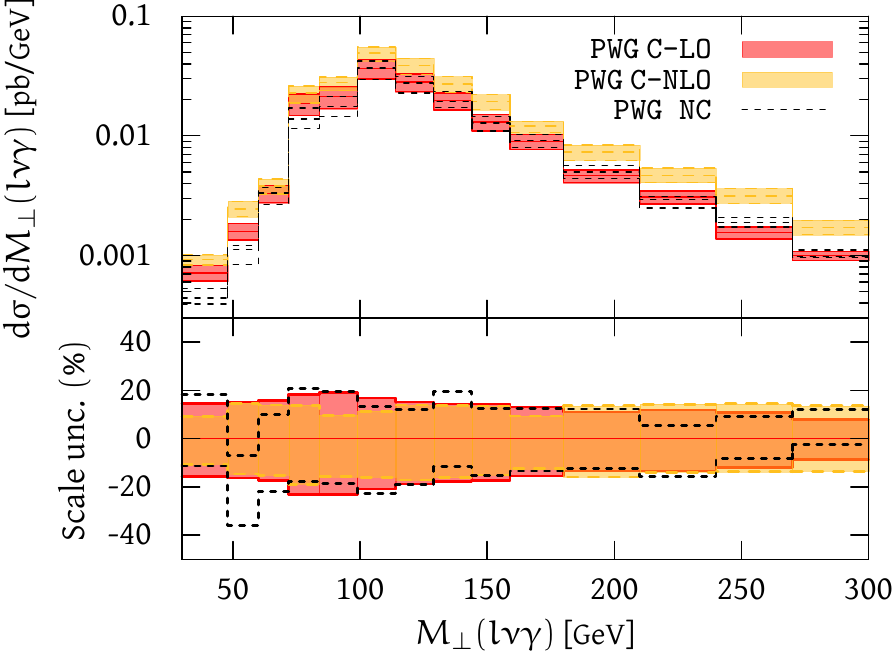}
\end{center}
\caption{\label{Scale} NLOPS simulations obtained with \POWHEG{}-NC, \POWHEG{}-C-LO and \POWHEG{}-C-NLO for the $\pt^{\gamma}$ 
and $M_T^{\ell \nu \gamma}$ distributions in the basic photon (upper plots) and lepton cuts (lower plots)
conditions, respectively. The bands are an estimate of the theoretical uncertainty derived from the scale variation.}
\end{figure}

Examples of NLOPS simulations of differential cross sections including an estimate of the theoretical uncertainty,
are given in Fig.~\ref{Scale}. 
We obtained the bands from the upper- and lower-bounding envelopes of the distributions, as already explained, 
but in terms of the variations of the dynamical factorization and renormalization scale described in Section~\ref{sec:mcfm}.
We just show the results for the  $\pt^\gamma$ 
and $M_T^{\ell \nu \gamma}$ distributions in the basic photon (upper plots) and lepton cuts (lower plots) conditions, and 
we present the predictions of all the \POWHEG{}+MiNLO implementations, {\it i.e.} \POWHEG{}-NC
 (dashed lines) \POWHEG{}-C-LO (orange band) 
and \POWHEG{}-C-NLO (light orange band). The latter realizations share the same $O(\as^2)$ 
$Wjj + \gamma_{\rm PS}$ dynamics for the 
description of the real radiation mechanism, but differ in the LO(NLO) accuracy in the calculation of the $Wj$ normalization.
It can be seen that our estimate produces rather large theoretical uncertainties, as 
already remarked for the integrated cross section results. Moreover, the $Wj(j)$ NLO normalization contributions are not
 irrelevant at all in the whole $\pt^\gamma$ 
and $M_T^{\ell \nu \gamma}$ range, especially in the extreme kinematical regions.
As discussed earlier, this means that the corrections beyond NLO accuracy can
play a relevant r\^ole for a reliable extraction of limits on ATGCs
from the high tails of the $\pt^\gamma$ differential cross section,
especially in view of the experimental accuracy expected at the Run II
of the LHC.

\subsection{Comparisons to LHC data at $\sqrt{s}$ = 7~TeV}
\label{sec:analysis}
In this Section we 
compare our predictions with all the data published by ATLAS collaboration for $W\gamma$ 
production at the LHC at $\sqrt{s}$ = 7~TeV~\cite{Aad:2013izg}.\footnote{We do not provide comparisons 
with CMS data, since they are quoted after corrections from MC simulations and just refer to 
three inclusive cross sections for $\pt^{\gamma, {\rm min}} = 15, 60, 90$~GeV and a photon-lepton 
separation $\Delta R(\ell, \gamma) > 0.7$~\cite{Chatrchyan:2013fya}.} The lepton acceptance cuts and photon isolation criteria 
applied by ATLAS collaboration are the same as in Eq.~(\ref{eq:cut3}), but jet identification requirements
are also applied as follows
\begin{eqnarray}
 E_T^{\rm jet} > 30~{\rm GeV}, \, |\eta_{\rm jet}| < 4.4, \, \Delta R (e/\mu/\gamma, {\rm jet}) > 0.3 \; .
\end{eqnarray}
Moreover, jets are defined according to an anti-$k_T$ recombination algorithm, with 
separattion parameter $R_0 = 0.4$.\footnote{We implemented the jet algorithm using the FastJet code~\cite{Cacciari:2011ma}.} 
The data quoted by ATLAS  refer to an inclusive ($N_{\rm jet} \ge 0$) and exclusive ($N_{\rm jet} = 0$) event selection.
We present results for both conditions. For simplicity, the measured 
cross sections are compared with the predictions of the most accurate NLOPS simulations
\POWHEG{}-C-LO and \POWHEG{}-C-NLO. 

\begin{figure}[t]
\begin{center}
\includegraphics[width=7.5cm]{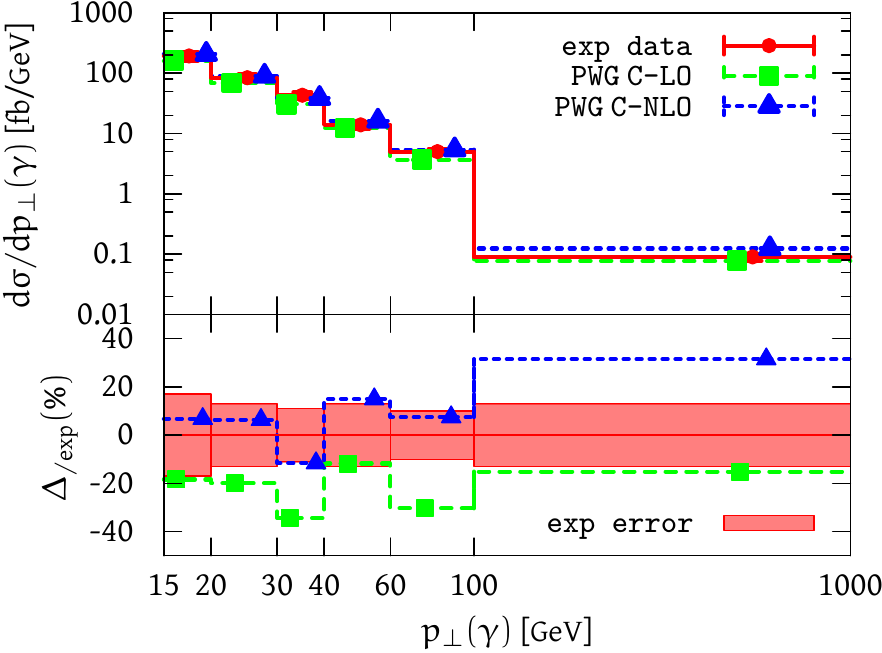}~~~~~\includegraphics[width=7.5cm]{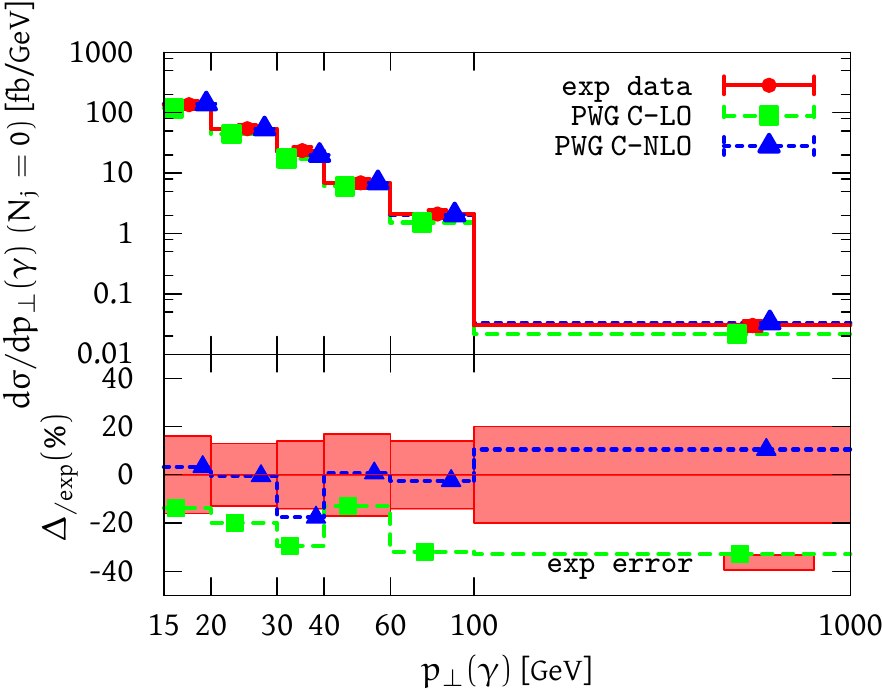}
\end{center}
\caption{\label{ptgATLAS} The $\pt^{\gamma}$ differential cross sections of the $p p \to \ell \nu \gamma$ process measured by the ATLAS 
collaboration at $\sqrt{s} = 7$~TeV in  the inclusive $N_{\rm jet} \ge 0$ (left) and exclusive $N_{\rm jet} = 0$ (right) event selections, in 
comparison to the NLOPS \POWHEG{}+MiNLO predictions. The lower panels show 
$\Delta / {\rm exp} = (d\sigma_{\rm th.} - d\sigma_{\rm exp.})/d\sigma_{\rm exp.}$, taking into 
account the total experimental uncertainty.}
\end{figure}

For the integrated cross section for the $\ell\nu\gamma$ process, $\ell = e^{\pm}$, we get the predictions
\begin{enumerate}
\item \POWHEG{}-C-LO \, \, \, $\sigma (N_{\rm jet} \ge 0)  {\rm [pb]} \, = \,  2.25^{+0.24}_{-0.24}$ \quad \, $\sigma (N_{\rm jet} = 0)  {\rm [pb]} \, = \, 1.42^{+0.15}_{-0.15}$
\item \POWHEG{}-C-NLO  \, \, $\sigma (N_{\rm jet} \ge 0)   {\rm [pb]} \, = \,  2.95^{+0.20}_{-0.38}$ \, \, \, $\sigma (N_{\rm jet} = 0)   {\rm [pb]} \, = \, 1.69^{+0.11}_{-0.22}$
\end{enumerate}
where the theoretical uncertainties have been estimated from renormalization and factorization scale variations, according to the procedure 
described in Section \ref{sec:mcfm}. These predictions must be compared to the ATLAS measured values~\cite{Aad:2013izg}
\begin{eqnarray*}
\sigma_{\rm exp}  (N_{\rm jet} \ge 0)  {\rm [pb]} \, = \,  2.74 \, {\pm} \, 0.05 {\, \rm (stat)}   \, {\pm} \, 0.32 {\, \rm (syst)}  \, {\pm} \, 0.14 {\, \rm (lumi)}  \\
\sigma_{\rm exp}  (N_{\rm jet} = 0)  {\rm [pb]} \, = \,  1.77 \, {\pm} \, 0.04 {\, \rm (stat)}   \, {\pm} \, 0.24 {\, \rm (syst)}  \, {\pm} \, 0.08 {\, \rm (lumi)}
\end{eqnarray*}
The \MCFM{} results quoted by ATLAS for these configurations, after parton-to-particle level corrections, are
\begin{eqnarray*}
\sigma_{\rm \MCFM{}}  (N_{\rm jet} \ge 0)  {\rm [pb]} \, = \,   1.96  \, {\pm} \, 0.17 \quad \quad \quad \sigma_{\rm \MCFM{}}  (N_{\rm jet} = 0)  {\rm [pb]} \, = \, 1.39 \, {\pm} \, 0.17
\end{eqnarray*}
where the errors are an estimate of the theoretical systematic uncertainties, obtained by varying the PDFs, by changing the definition of
photon isolation and by variation of the renormalization and factorizations scales from the nominal value $\mu_R = \mu_F = \sqrt{ M_W^2 + \pt^{\gamma \, 2} }$
up and down by a common factor of two.

It can be seen that our results, in particular the predictions provided by \POWHEG{}-C-NLO, are in good agreement with the measured 
integrated cross sections, for both the inclusive and exclusive event selections. 

\begin{figure}[t]
\begin{center}
\includegraphics[width=9cm]{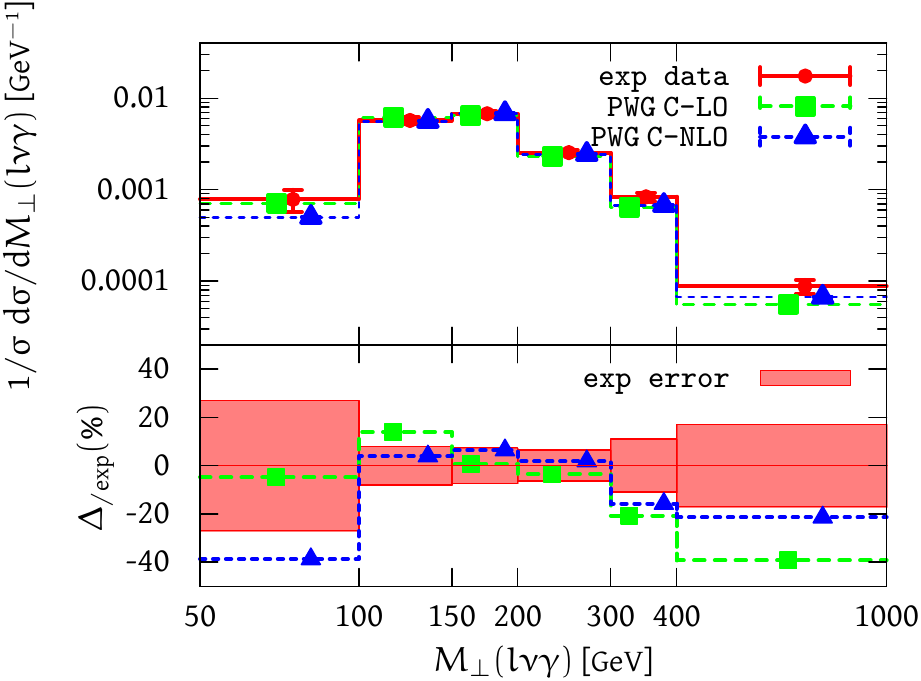}
\end{center}
\caption{\label{mtATLAS} The normalized differential cross section of the $p p \to \ell \nu \gamma$ process measured by the ATLAS 
collaboration at $\sqrt{s} = 7$~TeV in the inclusive $N_{\rm jet} \ge 0$ event selection as a function of $M_T^{\ell \nu \gamma}$ 
(for $\pt^{\gamma} >$~40 GeV), 
 in comparison to the NLOPS \POWHEG{}+MiNLO predictions. The lower panels show 
 $\Delta / {\rm exp} = (d\sigma_{\rm th.} - d\sigma_{\rm exp.})/d\sigma_{\rm exp.}$, taking into 
account the total experimental uncertainty.}
\end{figure}

\begin{figure}[t]
\begin{center}
\includegraphics[width=7.5cm]{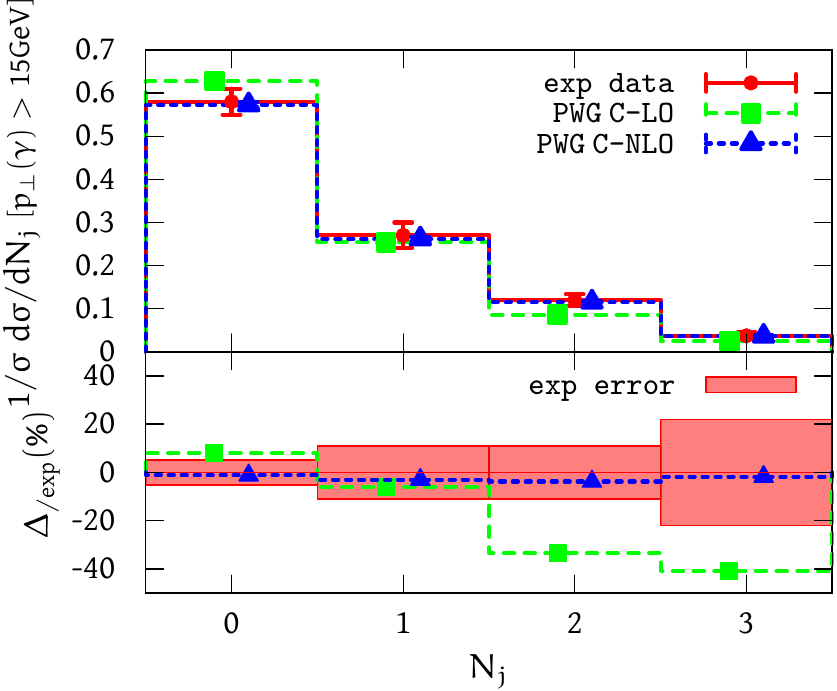}~~~~~\includegraphics[width=7.5cm]{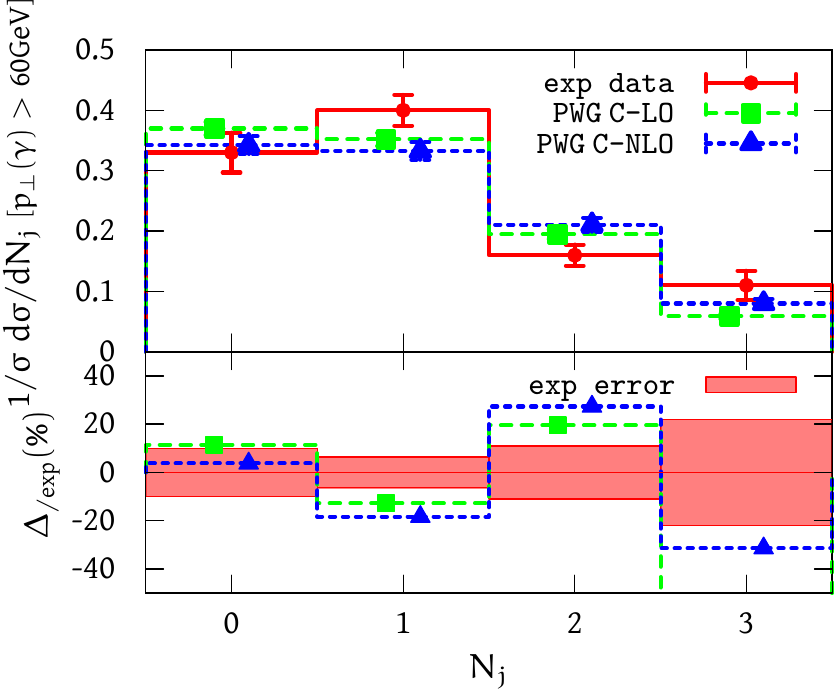}
\end{center}
\caption{\label{njetATLAS} The differential cross section measurements of ATLAS collaboration at $\sqrt{s} = 7$~TeV as 
a function of the jet multiplicity for the $p p \to \ell \nu \gamma$ process, for $\pt^{\gamma} >$~15~GeV (left) and 
$\pt^{\gamma} >$~60~GeV (right),  in comparison to the NLOPS \POWHEG{}+MiNLO predictions. The lower panels 
show $\Delta / {\rm exp} = (d\sigma_{\rm th.} - d\sigma_{\rm exp.})/d\sigma_{\rm exp.}$, taking into 
account the total experimental uncertainty.}
\end{figure}

The measured differential cross sections of the $p p \to \ell \nu \gamma$ process, obtained using combined 
electron and muon measurements,  are shown in comparison to the NLOPS \POWHEG{}+MiNLO 
simulations in Fig. \ref{ptgATLAS} for the  $\pt^{\gamma}$ distribution in the inclusive and exclusive event selection,
in Fig.~\ref{mtATLAS} for normalized cross section as a function of 
$M_T^{\ell \nu \gamma}$ in the inclusive case (under the condition $\pt^{\gamma} >$~40 GeV), and in Fig. \ref{njetATLAS} for the
normalized cross section as a function of the jet multiplicity with 
$\pt^{\gamma} >$~15~GeV  and 
$\pt^{\gamma} >$~60~GeV. In the lower panel of each plot we show the relative deviation 
$\Delta / {\rm exp} = (d\sigma_{\rm th.} - d\sigma_{\rm exp.})/d\sigma_{\rm exp.}$ of the 
predictions to the data,
including the total experimental uncertainty.

As can be seen, our theoretical results agree well with the data, as they reproduce the normalization and shape of the differential 
cross sections with good accuracy. In particular, while the predictions of \POWHEG{}-C-LO are slightly lower than the data, 
the results of \POWHEG{}-C-NLO, that includes the $Wj$ dynamics with normalization at NLO accuracy, are in 
very good agreement with the measurements.

\section{Conclusions}
\label{sec:conc}

In this work we have presented an application of the \POWHEG{} method, supplemented with
the MiNLO procedure, to the theoretical 
treatment and simulation of the $W(\ell\nu)\gamma$ production process in hadronic collisions. We have shown 
how the method can be modified to cope with both the direct photon production and fragmentation contribution. 
To this end, we have used the matrix elements associated with the two different $W\gamma$ and $Wj$ 
contributing processes and then 
showered and hadronized the generated events with a mixed QCD+QED parton shower,
in order to provide  a complete and fully exclusive description of the 
process under consideration. We have devised three different \POWHEG{}+MiNLO descriptions of
the $W\gamma$ process, characterized by increasing theoretical accuracy through the usage of 
QCD cross sections with LO/NLO accuracy matched to a PS. In particular, the method used in our most accurate 
simulations is a novel approach, that can be applied to any prompt-photon production process.

Our generator also includes the contribution of anomalous gauge couplings,
in order to provide a complete tool for data analysis at the
LHC. The code is available in the public repository of the
\POWHEGBOX{}~\cite{Alioli:2010xd} at the web site {\tt
http://powhegbox.mib.infn.it}.

In order to test the reliability of our calculation, we have compared
our results with the predictions of the fixed-order MC program
\MCFM{}.  In spite of the different treatment of the fragmentation
mechanism, as well as of the presence in our approach of higher-order
perturbative contributions, Sudakov and shower effects not included in
\MCFM{}, we observe an acceptable agreement with \MCFM{}. However,
this comparison also points out the relevance of higher-order perturbative and PS corrections
for a reliable modeling of the differential cross sections of
experimental interest, with a non-trivial dependence of those
contributions on the considered event selection condition.

An alternative NLOPS-based method for the simulation of the interleaved
QCD+QED emission off partons in a prompt-photon production processes
was proposed in ref.~\cite{D'Errico:2011sd}. We have also provided a
variant of our generators that mimics its behaviour, and have found
that this variant provides a good approximation of the QCD+QED
competition mechanism.

A yet different approach to direct photon production was used,
for example, in refs.~\cite{Kardos:2014zba,Kardos:2014pba}, to study $t\bar{t}\gamma$ and $t\bar{t}\gamma\gamma$
production. There, the photon emission is treated just like any other
hard process in \POWHEG{}, and the singularity associated to the
photon are screened by introducing a technical cut, consisting
in a smooth cone isolation applied to the photon. This is introduced
using rather loose isolation parameters, assuming that they will not
affect the cross section for realistic cuts.
This approach is extremely simple, since it does not require either
modifications to the \POWHEGBOX{} machinery or mixed QED+QCD showers.
 In light of our present study, we consider this approach viable if the radiated
photon is harder that the accompanying jets. If this is not the case,
especially for a photon softer than two other jets in the event ({\it i.e.}
when one jet is generated by the shower) the production probability does not
reflect a consistent approximation to the real dynamics.

More importantly, we have compared our NLOPS simulations with the data published by ATLAS 
collaboration for the $p p \to \ell \nu \gamma$ process at the LHC at 7 TeV. We observe a good
agreement between our predictions and the measured cross sections, for both the inclusive and exclusive 
event selections. In particular, both the normalization and shape of the measured differential cross section 
are reproduced remarkably well by our most accurate calculation.

The work presented here describes the first NLOPS simulation of the $W\gamma$ production process at hadron colliders. 
It contains several novel features in comparison to the existing theoretical literature about isolated 
photon hadroproduction and paves the way to the realization of future NNLOPS simulations 
of prompt photon production in a hadronic environment, thanks to the recent progress in 
this area~\cite{Hamilton:2013fea,Hoeche:2014aia,Karlberg:2014qua,Hoche:2014dla} and the 
calculation of NNLO QCD corrections to specific processes involving isolated 
photons~\cite{Grazzini:2013bna, Grazzini:2014pqa, Catani:2011qz}.
The approach can be easily adapted to deal with other relevant prompt-photon production reactions in hadronic collisions, 
such as $Z\gamma$ and $\gamma \gamma$ production. Moreover, the exclusive MC modeling of the 
fragmentation contribution presented in the paper provides a new way of performing interesting 
QCD studies of the measured photon fragmentation functions and of their typical 
theoretical parameterizations.

\acknowledgments

This work was supported in part by the Research Executive Agency (REA) of the European Union 
under the Grant Agreement number PITN-GA-2010- 264564 (LHCPhenoNet), and by the Italian 
Ministry of University and Research 
under the PRIN project 2010YJ2NYW. The work of L.B. is supported by the ERC grant 
291377, ``LHCtheory - Theoretical predictions and analyses of LHC physics: advancing the precision frontier". 
F.P. wishes to thank the CERN PH-TH Department for partial support and hospitality 
during several stages of the work. Useful correspondence with John Campbell on \MCFM{} and 
with Peter Skands about \PYTHIA{} is 
gratefully acknowledged. The authors thank the Galileo Galilei Institute for Theoretical Physics for 
hospitality and the INFN for partial support during the revision of this work.

\bibliographystyle{JHEP}
\bibliography{Wgamma}

\end{document}